\documentclass[journal]{IEEEtran}
\usepackage{amsfonts}
\IEEEoverridecommandlockouts

\ifCLASSINFOpdf
\else
\fi
\usepackage{subfigure}
\usepackage{epstopdf}
\usepackage{multirow}
\usepackage{cite}
\usepackage{stfloats}
\usepackage{color}
\usepackage{epsfig}
\usepackage{graphicx}
\usepackage{psfig}
\usepackage{epsf}
\usepackage{float}
\usepackage{amssymb }
\usepackage[cmex10]{amsmath}
\usepackage{algorithm} 
\usepackage{algorithmic}
\usepackage{booktabs}
\usepackage{amsmath,bm}
\usepackage{caption}

\usepackage[colorlinks,
            linkcolor=blue,       
            anchorcolor=blue,     
            citecolor=blue,       
            ]{hyperref}

\makeatletter
\renewcommand{\maketag@@@}[1]{\hbox{\m@th\normalsize\normalfont#1}}%
\makeatother

\usepackage{tikz,xcolor}

\definecolor{lime}{HTML}{A6CE39}
\DeclareRobustCommand{\orcidicon}{
\begin{tikzpicture}
\draw[lime, fill=lime] (0,0)
circle[radius=0.16]
node[white]{{\fontfamily{qag}\selectfont \tiny \.{I}D}};
\end{tikzpicture}
\hspace{-2mm}
}
\foreach \x in {A, ..., Z}{%
\expandafter\xdef\csname orcid\x\endcsname{\noexpand\href{https://orcid.org/\csname orcidauthor\x\endcsname}{\noexpand\orcidicon}}
}

\begin{document}
\title{Symbiotic Backscatter Communication: A Design Perspective on the Modulation Scheme of Backscatter Devices}
\author{Yinghui Ye, Shuang Lu, Liqin Shi, Xiaoli Chu, and Sumei Sun,~\emph{Fellow, IEEE}
 \thanks{ Yinghui Ye, Shuang Lu and Liqin Shi are with the Shaanxi Key Laboratory of Information Communication Network and Security, Xi'an University of Posts $\&$ Telecommunications, China. {\emph{(Corresponding author: Liqin Shi.)}} }
 \thanks{Xiaoli Chu is with the School of Electrical and Electronic Engineering,
University of Sheffield, Sheffield, U.K. } 
 \thanks{Sumei Sun is  is with the Institute of Infocomm Research, Agency for Science,
Technology and Research, Singapore. } 
}
\markboth{}
{Shi\MakeLowercase{\textit{et al.}}:}
\maketitle

\begin{abstract}
   Symbiotic Backscatter Communication  (SBC) has emerged as a spectrum-efficient and low-power communication technology, where backscatter devices (BDs) modulate and reflect incident  radio frequency (RF) signals from primary transmitters (PTs). While previous studies have  assumed a circularly symmetric complex Gaussian (CSCG) distribution for the BD's signal, this assumption may not be practical because the high complexity of generating CSCG signals is not supported by the low-cost BD. In this paper, we address this gap by investigating SBC for two low-complexity modulation schemes, i.e.,  $M$-ary amplitude-shift keying (MASK) and $M$-ary phase-shift keying (MPSK), where BD's signals inherently deviate from CSCG distribution. Our goal is to derive the achievable rate of the PT and BD under  the MASK/MPSK  and to design MASK/MPSK modulation scheme for maximizing the PT's rate.
   Towards this end, we first derive the expressions of both the PT's rate and BD's rate.
   Theoretical results reveal that whether or not the BD improves the PT's rate depends on the phase of MASK/MPSK modulation, while the BD's rate is independent of this phase. We then formulate two optimization problems to maximize the PT's rate by adjusting the phase under the  MASK and MPSK modulation schemes, respectively, and derive the optimal phases for each modulation scheme  in closed forms. {\color{black} Given that the  optimal phase is continuous and thus impractical for real-world BDs, we also propose a practical circuit design that enables BDs to select a  discrete phase close to the  theoretical optimum.}
   Simulation results demonstrate that the optimal phase of MASK/MPSK can ensure an improvement in the PT's rate and {\color{black}the performance gain between the ideal continuous phase and the practical implementation with few discrete phases is negligible}, and reveal that a low-order  ASK modulation is better than a low-order PSK for the BD in terms of improving PT's rate,   especially when the direct link is not significantly weaker than the backscatter link in SBC.
\end{abstract}

\IEEEpeerreviewmaketitle

\section{Introduction}
{\color{black}By 2030, the number of Internet of Things (IoT) devices worldwide is projected to exceed 80 billion \cite{8879484}, imposing significant pressure on wireless connectivity due to scarce radio spectrum resources and rapidly growing energy demands. IoT devices can be broadly classified into four categories based on application requirements: broadband IoT, critical IoT, massive IoT, and ambient IoT \cite{10463656}. Among these, ambient IoT represents the  most cost- and energy-constrained class of applications, characterized by requirements for ultra-low device complexity and ultra-low power consumption. Recognizing its potential, the 3rd Generation Partnership Project (3GPP) initiated standardization efforts in 2022 and identified backscatter communication as a key enabling technology for ambient IoT \cite{11015785}.}

In this context, Symbiotic Backscatter Communication (SBC) has emerged as a pivotal research direction in backscatter
communications, enabling BDs to harness ambient RF signals for information transmission and thus achieving a spectrum- and energy-efficiency technology \cite{9749195, xu2025revolutionizing,9193946}. It allows backscatter devices (BDs) to convey information by modulating and reflecting the incident   radio frequency (RF) signals transmitted by primary transmitters  (PTs). This approach eliminates the need for additional spectrum and carrier signal generation by the BD, thus  enabling  spectrum-efficient and low-power information transmission \cite{9051982,10980384}.

The concept of SBC was introduced in \cite{8907447}.  Assuming  the circularly symmetric complex Gaussian
(CSCG) signal for the BD and the significantly longer symbol duration\footnote{There have been several works assuming that the BD's symbol duration is comparable to that of the PT (see \cite{10643608, 10964549, 10502325} and reference therein). However, in this work, we focus on the case where the BD's symbol duration is much longer to that of the PT, thus these works  \cite{10643608, 10964549, 10502325} have not been reviewed.} of the BD compared to the PT, the authors demonstrated that the BD's signal could effectively enhance the PT's rate \cite{8907447}. This makes SBC particularly appealing in the era of IoT for the following reasons. In traditional spectrum-sharing communications, the interference from spectrum-sharing transmitters typically degrades the PT's performance. However, in SBC, it has been shown that when the BD's signal is properly utilized, the interference from the BD vanishes and can even contribute to improving the PT's transmission performance. This presents a novel and promising spectrum-sharing paradigm, encouraging numerous contributions to optimize and evaluate the performance of SBC.

The authors in \cite{8907447} maximized the weighted sum rate of both the PT and BD by jointly optimizing the PT's transmit power and beamforming vectors in SBC system.
In \cite{9461158}, the energy efficiency of a SBC network was maximized by jointly optimizing the transmit power of the PT, the reflection coefficients and backscattering time of the BD.
Considering the hardware impairments at transceivers, the authors in  \cite{9866050} maximized the weighted sum rate  of a  BD and multiple BDs in the SBC network.
In \cite{8665892}, the authors maximized the weighted sum rate of the PT and BD by jointly optimizing the PT's transmit power and the BD's reflection coefficient under either long-term or short-term transmit-power constraint over the fading channel. Although many resource allocation schemes have been proposed to enhance the BD's transmission, the improvement is still limited due to the large difference in symbol duration  between PT and BD.
To address this issue and ensure the improvement of PT's rate, the author in \cite{10437703} proposed a novel hybrid active-passive SBC, where the BD transmits information via  passive backscatter communication and active communication alternatively, and maximized the sum rate of all BDs, while ensuring that the PT's rate is larger than that without the access of BD.
The time allocation, BD's reflection coefficient and active communication transmit power were jointly optimized to minimize the total transmission time of all BDs in the hybrid active-passive communication \cite{10820118}.

Beyond resource allocation \cite{8907447, 9461158, 9866050, 8665892, 10437703, 10820118}, the performance evaluation in the SBC network was also investigated.
The authors in \cite{8807353} derived the upper bounds of the ergodic capacity for both the PT and BD links.
In \cite{8941106}, the authors derived the optimal reflection coefficient of BD and the optimal transmit power of PT, and then obtained closed-form expressions of the outage probability for both the PT and BD.
In contrast to \cite{8807353} and \cite{8941106}, which considered only a  BD, the authors in  \cite{10702412} proposed the random selection access  scheme and the selection diversity access scheme for the case of multiple BDs, and analyzed the outage performances of both the PT and BDs under these schemes. Considering the  possibility that the link between BD and its associated receiver may be blocked, the authors in \cite{10778600} proposed a novel relay-assisted SBC, and analyzed the outage performance of PT and BD under the three forwarding schemes. The authors in \cite{10896822} employed a reconfigurable intelligent surface to relay signals from  both the PT and BD, and  analyzed the corresponding outage probabilities.

We note that  the existing works \cite{8907447,9461158,9866050,8665892,10437703,10820118,8807353,8941106,10702412,10778600,10896822} assumed  a CSCG distribution for the BD reflected symbols.  However, this assumption may be unrealistic in practical SBC for the following reasons. While a CSCG  signal  maximizes the mutual information between the BD and its receiver, it necessitates  high-complexity techniques such as probabilistic shaping \cite{8006941} to transform signals after amplitude or phase modulation into CSCG signals. These techniques, however, are not supported by the low-cost  BD, which typically relies on relatively simple circuits \cite{1505023,7059230,10681516}.

In this paper, we focus on  the SBC, where the BD\footnote{{\color{black}A reconfigurable intelligent surface (RIS) is typically designed to enhance wireless transmission performance by optimizing channel conditions, whereas a backscatter device (BD) is aimed at delivering information with ultra-low
device complexity and ultra-low power consumption. It is worth noting that recent research has explored the possibility of enabling RIS to modulate information \cite{9309091}, suggesting that an RIS could, in principle, replace the BD in our considered system model. However, the hardware complexity and manufacturing cost of RIS units are substantially higher than those of BDs. Therefore, RIS may not be a suitable solution for large-scale deployments  of ultra low-power and low-cost IoT devices. This practical constraint motivates our focus on the use of BDs in this work.} } adopts one of the following two popular low-complexity modulation schemes, i.e.,  $M$-ary amplitude-shift keying (MASK) and $M$-ary phase-shift keying (MPSK)\footnote{\color{black}MASK and MPSK are selected for their dual advantages of being recommended by 3GPP for IoT and enabling simple, low-power passive implementations suitable for BDs \cite{11015785}.}, to convey information. {\color{black}It should be noted that under MASK/MPSK, the BD's signal does not follow a CSCG distribution. Consequently, the following questions arise:
{\emph{First,  can BD still guarantee an increase in the PT's rate when using MASK or MPSK?
If not, how should the phase of the BD's symbols be designed to ensure an improvement in the PT's rate?}}
{\emph{Second, the symbols generated by the modulation scheme do not follow CSCG distribution, and applying Shannon formula would lead to an overestimation of the achievable rate.
In this case, what is the appropriate method for calculating the BD's rate?}}  Although  the recent work \cite{11020593} employs low-complexity modulation schemes, it does not address the above questions. Our work aims to fill this gap by providing a practical and theoretically grounded design framework for SRBC under realistic modulation constraints.}

Our main contributions are listed below:

{\color{black}\begin{itemize}
 \item Considering the  MASK or MPSK modulation scheme  employed by the BD, we derive the achievable rates of the PT and BD. The PT's rate is also obtained under the  infinite-order ASK and PSK, respectively.
 \item We theoretically prove the following three results.
     First, under the finite-order ASK or PSK modulation, whether or not allowing BD to backscatter information benefits PT's rate depends on the phase of the BD's symbols.
     This highlights the importance of optimizing the phases of  MASK and MPSK modulation schemes  in  practical SBC.
     Second, we demonstrate how the phase of the information transmitted by BD affects PT's transmission rate as the modulation order $M$ approaches to infinity under ASK and PSK modulation, respectively. Specifically, for the infinite-order PSK, the improvement in the PT's rate can  always  be ensured, while such a conclusion does not hold for the infinite-order ASK.
     Third, the rate of BD is independent of the phase of  the BD symbols.
   \item We formulate two  problems to maximize the PT's transmission rate by optimizing the phase of the BD's reflected signals under MASK and MPSK modulation schemes, respectively.
       We derive the closed-form expressions of the optimal phase for maximizing  PT's transmission rate under MASK and BPSK modulation, respectively.
       Our results demonstrate that for MASK, the optimal phase is determined solely by the channel coefficients and is independent of the modulation order.
       In contrast, for MPSK, the optimal phase depends on both the channel coefficients and the modulation order.
 To bridge theory and practice, we further propose a practical circuit design that enables BDs to realize discrete phases, closely approximating the optimal performance.
       \item Simulation results  confirm that the optimal phase of MASK/MPSK can improve the PT's rate, and  demonstrate that the proposed discrete-phase circuit design achieves near-optimal performance. Besides, the results  also reveal the condition under which ASK outperforms PSK in terms of the {\color{black}PT's rate.}
\end{itemize}}

%


This paper is organized as follows.
Section II describes the SBC system model.
In Section III, we derive the achievable rates for the PT and BD, targeting the adoption of MASK, and MPSK modulation, respectively.
Section IV formulates the PT's rate maximization problem under both MASK and MPSK modulation schemes and derives the optimal phase.
Section V presents numerical results.
Finally, the paper is concluded in Section VI.

{\emph{Notations:}} $X^*$, $|X|$, ${\rm{Re}}(X)$, and ${\rm{arg}}(X)$ denote the conjugate, the amplitude, the real part, and the  argument of a complex number $X$, respectively.  $j$ is the imaginary unit.
\section{System Model}
\begin{figure}
\centering
  \includegraphics[width=0.35\textwidth]{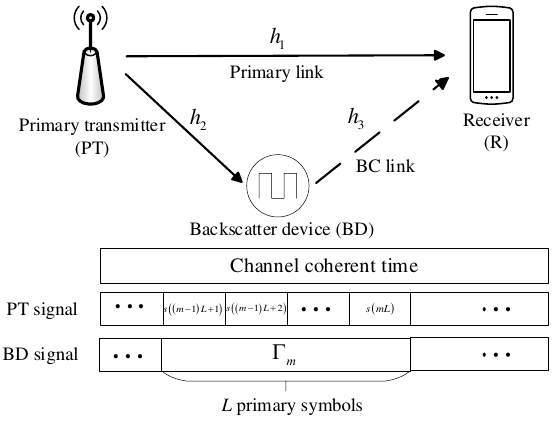}\\
  \caption{System model and time scheduling structure.}\label{fig1}
\end{figure}

Fig. \ref{fig1} depicts a  SBC system, where the receiver (R) simultaneously  extracts the PT's information and the BD's information that is modulated onto the PT's signal. Let $h_1$, $h_2$,  and $h_3$ denote the channel coefficients of the PT-R link, the PT-BD link, and the BD-R link, respectively, where $|h_i|$ and $\theta_i$ $\left(i=1,2,3\right)$ represent the amplitude and the phase of $h_i$,  respectively.
All the channel coefficients are assumed to  remain fixed within a  transmission block but may change across different blocks.
Let $s\left(n\right)$ denote the signal transmitted by the PT, which has a mean of zero
and a variance of one, with a symbol period  of  $T_s$.
The period of a  BD symbol is denoted by $T_c$. Since the period of a  BD symbol is much longer than that of the PT,  we assume  $T_c=LT_s$, where  $L\gg1$ is a positive integer \cite{8907447,9461158,9866050,8665892,10437703,10820118,8807353,8941106,10702412,10778600,10896822}.

\begin{figure}
\centering
  \includegraphics[width=0.35\textwidth]{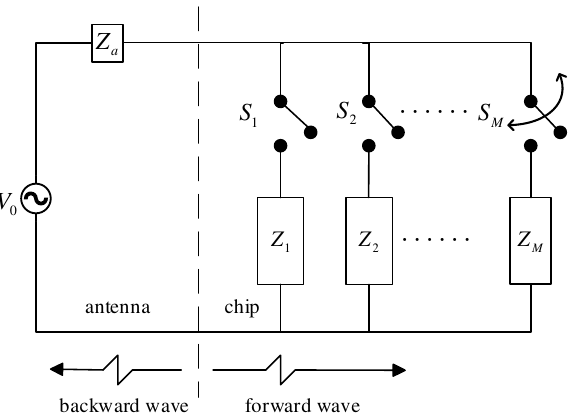}\\
  \caption{BD circuit structure.}\label{fig2}
\end{figure}

In SBC, the BD transmits equidistant and equiprobable symbols\footnote{Considering equiprobable symbols at the BD is a practical choice and the reason is summarized as follows. Due to its hardware simplicity and energy constraint, the BD requires a statistically efficient encoding scheme to minimize the use of symbol resources, which motivates us to consider equiprobable symbols at the BD since it maximizes entropy  from an information-theoretic perspective. While techniques like probabilistic shaping could theoretically enhance mutual information by adapting the distribution of  symbols modulated by the BD, their computational overhead conflicts with the BD's inherent need for low-complexity operation.}, corresponding to an $M$-order symbol sequence $\left\{ {c\left( m \right),  m = 1,2, \ldots M } \right\}$. The $M$ distinct  symbols are mapped to $M$ complex reflection coefficients $\left\{ {{\Gamma _m},m = 1,2, \ldots M} \right\}$ through an appropriate adjustment of load impedances  $\left\{ {{Z_m}, m = 1,2, \ldots M} \right\}$ \cite{1233745, 7820135}, as shown in Fig. \ref{fig2}.
Without loss of generality, we assume that the $m$-th  symbol of the BD is uniquely mapped to the $m$-th complex reflection coefficient, achieved by  adjusting the  $m$-th load impedance.
Thus, the backscattered signal of the BD is written as ${\sqrt P {h_2}{\Gamma _m}s\left( n \right)}$.
 Here, ${\Gamma _m}$ keeps unchanged for $ n = \left( {m - 1} \right)L + 1,\left( {m - 1} \right)L + 2,...,mL$, due to  $T_c=LT_s$, and   is  calculated as  \cite{6153042}
\begin{align}\label{1}
{\Gamma _m} = \frac{\left(Z_a\right)^* - {Z_m}}{{{Z_a} + {Z_m}}}\mathop  = \limits^\Delta  {\alpha _m}\exp \left( {j{\varphi _m}} \right),
\end{align}
where  $Z_a=R_a+jX_a$ is the antenna impedance with the resistance $R_a$ and the reactance $X_a$,  $Z_m=R_m+jX_m$  denotes the load impedance with the load resistance $R_m$ and the reactance $X_m$, and $\alpha_m=\left| {{\Gamma _m}} \right|$ and  $\varphi _m={\rm{arg}}\left( {{\Gamma _m}} \right)$ are the amplitude  and the phase, respectively, corresponding to ${\Gamma _m}$.  Due to the constraints imposed by the  impedance values of passive components, the $M$ complex reflection coefficients  are confined to the complex plane within a circle centered at the origin, with a radius not exceeding one. Thus, we have  $ 0\le \alpha_m \le 1$ and $0\le \varphi _m\le 2\pi$.

The received signal at  R is expressed as
\begin{align}\label{2}
 y\left( n\right)&= \underbrace {\sqrt P {h_1}s\left( n \right)}_{{\rm{PT}} \to {\rm{R}}\;{\rm{link}}} + \underbrace {\sqrt P {h_2}{h_3}{\Gamma _m}s\left( n \right)}_{{\rm{PT}} \to {\rm{BD}} \to {\rm{R}}\;{\rm{link}}} + \omega \left( n \right) \nonumber\\
&=\sqrt {P}  h_{{\rm{eq}},m}s\left(n\right)+ \omega \left( n\right) ,
\end{align}
where  $h_{{\rm{eq}},m}\overset{\triangle}{=}  {{h_1} + {h_2}{h_3}{\Gamma _m}} $ keeps unchanged for $ n = \left( {m - 1} \right)L + 1,\left( {m - 1} \right)L + 2,...,mL$, and $\omega \left( n\right)$ is the additive white CSCG noise with zero mean and variance $\sigma ^{2}$.

Since $s(n)$ is a CSCG signal, we can apply   Shannon formula to calculate the PT's rate,  given a specific  BD's symbol, as ${\log _2}\left( {1 + \frac{{P|{h_{{\rm{eq}},m}}{|^2}}}{{{\sigma ^2}}}} \right)$. Then, using the Law of Total Probability, the PT's rate, in a  transmission block,  can be expressed as
\begin{align}\label{3}
{R_{\rm s}} = &\sum\limits_{m = 1}^M {p\left( {{\Gamma _m}} \right)} {\log _2}\left( {1 + \frac{{P{{| {{h_{{\rm{eq}},m}}} |}^2}}}{{{\sigma ^2}}}} \right),
\end{align}
where $p\left( {{\Gamma _m}} \right) = \frac{1}{M}$ is the probability
of occurrence of  $\Gamma_m$.

After successfully decoding $s\left(n\right)$,  the successive interference cancellation (SIC) technique is used at  R to decode  BD's signal.
Assuming that the PT's signal $s\left(n\right)$ can be perfectly removed from $y\left(n\right)$, then, the residual  signal can be written as
\begin{align}\label{4}
{\hat y\left( n \right)}= \sqrt P {h_2}{h_3}{\Gamma _m}s\left( n \right) + \omega \left( n \right).
\end{align}

Next,  R extracts BD's information from the residual signal based on \eqref{4}. As $\Gamma_m$ remains unchanged for  $ n = \left( {m - 1} \right)L + 1,\left( {m - 1} \right)L + 2,...,mL$, the residual signal can be viewed as  the BD's signal passing through $L$ wireless channels, i.e., $\sqrt P {h_2}{h_3}s\left( n \right)$. In this case, following \cite{8907447}, we assume that R uses  maximal ratio combining (MRC) on the residual, yielding
\begin{align}\label{5} \notag
{y_{\rm{MRC}}}\left( m \right)& = \frac{{{{\sum\limits_{n = 1}^L {\left( {\sqrt P {h_2}{h_3}s\left( n \right)} \right)} }^*}{\hat y\left( n \right)}  }}{{{\sigma ^2}}}\\
&= {g}{\Gamma _m} + {\omega _s},
\end{align}
where ${g} =\frac{{LP{{\left| {{h_2}} \right|}^2}{{\left| {{h_3}} \right|}^2}}}{{{\sigma ^2}}}$, and ${\omega _s}\sim \mathcal{CN} \left(0,\sigma _s^2\right)$ with $\sigma _s^2=\frac{{LP{{\left| {{h_2}} \right|}^2}{{\left| {{h_3}} \right|}^2}}}{{{\sigma ^2}}}$.



Due to the discrete symbol of the BD, 
using the Shannon formula would lead to an overestimation of its rate. {Instead, we apply the mutual information\footnote{From an information-theoretic perspective, the rate is typically characterized by the maximum  mutual information  by optimizing the distribution of the transmitted symbols. Specifically,  the BD's rate is expressed as $\mathop {\max }\limits_{p\left( \Gamma  \right)} I\left( {\Gamma ;{Y_{{\rm{MRC}}}}} \right)$, where ${p\left( \Gamma  \right)}$ is the probability of the BD's symbols. Since  the discrete uniform distribution is assumed for the BD's symbol, the maximization term is removed from  in \eqref{6}.} \cite{2011Mutual} to accurately characterize the BD's rate  as follows, }
\begin{align}\label{6}
R_{\rm{BD}} = I\left( {\Gamma ;{Y_{\rm{MRC}}}} \right),
\end{align}
where
\begin{align}\label{6a}  \notag
I&\left( {\Gamma ;{Y_{{\rm{MRC}}}}} \right) = \sum_{m = 1}^M \int_ {\mathbb{C} } {{\rm{Pr}}[\Gamma  = {\Gamma _m},{Y_{{\rm{MRC}}}} = {y_{\rm{MRC}}}]} \\
&\times {{\log }_2}\left( {\frac{{{\rm{Pr}}[\Gamma  = {\Gamma _m},{Y_{{\rm{MRC}}}} = {y_{{\rm{MRC}}}}]}}{{{\rm{Pr}}[\Gamma  = {\Gamma _m}] \cdot {\rm{Pr}}\left[ {{Y_{{\rm{MRC}}}} = {y_{{\rm{MRC}}}}} \right]}}} \right)d{y_{{\rm{MRC}}}}.
\end{align}
In \eqref{6a}, $\mathbb{C}$  represents the field of complex numbers, ${\rm{Pr}}\left[ \Gamma  = {\Gamma _m} \right]$ denotes the probability mass function at ${\Gamma _m}$ for a discrete random variable $\Gamma$, and ${\rm{Pr}}\left[ {Y_{{\rm{MRC}}}} = {y_{{\rm{MRC}}}} \right]$ represents the value of the probability density function at ${y_{{\rm{MRC}}}}$ for a continuous random variable ${Y_{{\rm{MRC}}}}$.

\section{Rate Analysis}
In this section, we analyze the rates of both  PT and BD under MASK  and MPSK modulation schemes, respectively.
If the BD adopts MASK modulation, the phase ${{\varphi _m}}$ remains unchanged, denoted by ${{\varphi _0^{\rm A}}}$, while variations in the amplitude $\alpha_m$ represent different information.
Conversely, if MPSK modulation is used, the amplitude $\alpha_m$ stays constant, denoted by $\alpha_0^{\rm{P}}$, and the information is encoded through variations in the phase  ${{\varphi _m}}$.

\subsection{ MASK Modulation}
In the case of BD using MASK modulation, there are $M$ distinct  symbols and  the $m$-th symbol is mapped into  ${\Gamma _{m }}$, given by ${\Gamma _{m}} = \frac{{m  - 1}}{{M - 1}}\exp \left(j{\varphi _0^{\rm A}}\right)$.
In this case, using \eqref{3}, the PT's  rate  can be rewritten as
\begin{align}\label{8}
&R_{\rm{s}}^{{\rm{A}}}\!=\! \frac{1}{M}\sum\limits_{m = 1}^M {{{\log }_2}\!\left( \!{1\! +\! \frac{{P|{h_1} \!+\! {h_2}{h_3}\frac{{m - 1}}{{M - 1}}\exp \!\left(\! {j{\varphi _0^{\rm A}}} \!\right)\!{|^2}}}{{{\sigma ^2}}}} \right)}.
\end{align}

One contribution of this work is to reveal the impact of the BD's modulation scheme on the PT's rate. To this end, we provide Lemma 1.

\textbf{Lemma  1:} If $\theta_2+\theta_3+\varphi _0^{\rm A} - \theta_1 =  \pm \pi $, and $\left| {{h_1}} \right|>\left| {{h_2}} \right|\left| {{h_3}} \right|$, then the rate gain of the PT can be expressed as
\begin{align}
\Delta {R_{\rm s}^{\rm{A}}} &= \sum\limits_{m = 1}^M {\frac{1}{M}{{\log }_2}\left( {1 + \frac{{P{{\left( {\left| {{h_1}} \right| - \frac{{m - 1}}{{M - 1}}\left| {{h_2}} \right|\left| {{h_3}} \right|} \right)}^2}}}{{{\sigma ^2}}}} \right)}  \nonumber \\
&- {R_p} < 0,
\end{align}
where ${R_p} = {\log _2}\left( {1 + \frac{{P{{| {{h_1}} |}^2}}}{{{\sigma ^2}}}} \right)$ denotes the PT's rate without BD access.

While if $\theta_2+\theta_3+\varphi _0^{\rm A} - \theta_1 =  0$, $\Delta {R_{\rm s}^{\rm{A}}}$  is given by
\begin{align}
\Delta {R_{\rm s}^{\rm{A}}} &= \sum\limits_{m = 1}^M {\frac{1}{M}{{\log }_2}\left( {1 + \frac{{P{{\left( {\left| {{h_1}} \right| +\frac{{m - 1}}{{M - 1}}\left| {{h_2}} \right|\left| {{h_3}} \right|} \right)}^2}}}{{{\sigma ^2}}}} \right)}  \nonumber \\
&-{R_p} > 0.
\end{align}


{\emph{Remark 1.}}  Lemma 1 indicates that whether  allowing BD to backscatter information benefits the PT's rate or not depends on the phase of BD's symbols under the MASK modulation.
This contradicts the well-known conclusion drawn in \cite{8907447}, where it is stated that allowing the BD to access the PT's spectrum can enhance the PT's rate.
This discrepancy arises from the distribution of  BD's symbols. Specifically, in \cite{8907447}, the BD's symbol is assumed to follow  a CSCG distribution, while in this work, we consider equidistant and equiprobable MASK symbols at BD.
Under the assumption of CSCG symbols, the phase of the BD's symbol is continuous  uniformly distributed from 0 to 2$\pi$, which is the key factor enabling the transformation of the BD's symbol into beneficial multipath components for the  primary transmission, thereby boosting the PT's rate.
However, under the MASK modulation, the phase of the BD's symbol keeps unchanged.
In this case, it is possible for  $|h_{{\rm{eq}},m}|<| {h_1} | $  to hold within  a  transmission block,  leading to a lower PT's rate compared to the scenario where BD access is not available.
Accordingly, in practical SBC with MASK modulation,  carefully designing the phase of  BD's symbols based on the phases of $h_1$, $h_2$, and $h_3$ is a prerequisite for transforming the BD's symbol into beneficial  components for the primary transmission.
Furthermore, for the design of BD, it is desirable to pre-establish a series of load  impedances that offer different phases under a given amplitude of the complex reflection coefficient.
This phase design concept is quite different from the conventional modulation scheme, where the phase of MASK remains fixed no matter what the the phases of $h_1$, $h_2$, and $h_3$  are.

In what follows, we derive the PT's rate by assuming $M \to \infty $.
Under this assumption, $\alpha_m$ follows the continuous uniform distribution  from 0 to 1, and the PT's rate can be calculated as \eqref{8b}, as shown at the top of the next page, where  ${C_1} = \frac{P}{{{\sigma ^2}}}{\left| {{h_1}} \right|^2} + 1$, $C_2=2\frac{P}{{{\sigma ^2}}}\left| {{h_1}} \right|\left| {{h_2}} \right|\left| {{h_3}} \right|\cos\left( {\theta_0 + \varphi _0^{\rm{A}}} \right)$ and $C_3=\frac{P}{{{\sigma ^2}}}{\left| {{h_2}} \right|^2}{\left| {{h_3}} \right|^2}$, $\theta_0=\theta_2+\theta_3- \theta_1$.
In \eqref{8b}, the second equality holds from the integration by parts, and the  last equality  is derived by using $\int \frac{{\left( {{N_1}x + {N_2}} \right)dx}}{{E + 2Fx + G{x^2}}} = {\color{black}\frac{N_1}{{2G}}}\ln \left| {E + 2Fx + G{x^2}} \right| + \frac{{{N_2}G - {N_1}F}}{{G\sqrt {EG - {F^2}} }}\arctan \frac{{Gx + F}}{{\sqrt {EG - {F^2}} }}$ when $EG > {F^2}$ \cite[eq. (2.103.5)]{zwillinger2007table},  {\color{black} and ${C_1}{C_3} - {\left( {\frac{{{C_2}}}{2}} \right)^2} = \frac{{{P^2}}}{{{\sigma ^4}}}{\left| {{h_1}} \right|^2}{\left| {{h_2}} \right|^2}{\left| {{h_3}} \right|^2}\left( {1 - \cos^2 \left( \theta_0 + \varphi _0^{\rm{A}} \right)} \right) + \frac{P}{{{\sigma ^2}}}{\left| {{h_2}} \right|^2}{\left| {{h_3}} \right|^2} > 0$.}

\begin{figure*}[ht] 
 	\centering
    \begin{align} \label{8b}
R_{\rm{s}}^{\rm{A}} &= \int_0^1 {{{\log }_2}\left( {1 + \frac{{P|{h_1} + {h_2}{h_3}\alpha \exp \left( {j\varphi _0^{\rm{A}}} \right){|^2}}}{{{\sigma ^2}}}} \right)} d\alpha \nonumber \\
  &= {\log _2}\left( {{C_1} + {C_2} + {C_3}} \right) - \frac{1}{{\ln 2}}\left( {\int_0^1 2 d\alpha  - \int_0^1 {\frac{{{C_2}\alpha  + 2{C_1}}}{{{C_1} + {C_2}\alpha  + {C_3}{\alpha ^2}}}} d\alpha } \right)\nonumber\\
  &{\color{black}={\log _2}\left( {{C_1} + {C_2} + {C_3}} \right) - \frac{2}{{\ln 2}} - \frac{{{C_2}}}{{2{C_3}\ln 2}}\ln \frac{{\left| {{C_1}} \right|}}{{\left| {{C_1} + {C_2} + {C_3}} \right|}} - \frac{{2{C_3}{C_1} - \frac{{{{\left( {{C_2}} \right)}^2}}}{2}}}{{{C_3}\sqrt {{C_3}{C_1} - {{\left( {\frac{{{C_2}}}{2}} \right)}^2}} \ln 2}}}\nonumber\\
  & \times {\color{black}\left( {\arctan \frac{{\frac{{{C_2}}}{2}}}{{\sqrt {{C_3}{C_1} - {{\left( {\frac{{{C_2}}}{2}} \right)}^2}} }} - \arctan \frac{{{C_3} + \frac{{{C_2}}}{2}}}{{\sqrt {{C_3}{C_1} - {{\left( {\frac{{{C_2}}}{2}} \right)}^2}} }}} \right)}
  \end{align}
  \hrulefill
\end{figure*}


Based on \eqref{8b}, we can obtain the following Lemma.

\textbf{Lemma 2:} For BD using MASK modulation, as $M \to \infty $, the PT's rate in the presence of BD  may be lower than that without BD access.
However, by carefully selecting the BD's modulation  phase, the BD's signal can bring an increase in the PT rate.

\emph{Proof:} Please refer to Appendix A.  \hfill {$\blacksquare $}

{\emph{Remark 2.}} The conclusion derived from Lemma 2 is identical to that of Lemma 1.
This is because, when BD uses MASK modulation,  the information is mapped to the amplitude of $\Gamma _{m}$ rather than its phase.

Next, we derive the BD's rate under MASK. For a given  specific
BD's symbol, it can be derived from \eqref{5} that ${y_{\rm{MRC}}}\left( m \right)\sim {\cal C}{\cal N}\left( {g{\Gamma _m},\sigma _s^2} \right)$. Using it and
 substituting ${\Gamma _{m }} = \frac{{m  - 1}}{{M - 1}}\exp \left(j{\varphi _0^{\rm A}}\right)$ into \eqref{6a}, we derive the BD's rate in  \eqref{9}, as shown at the top of the next page.

\textbf{Lemma 3:} $ R_{\rm{BD}}^{\rm{A}} $ is unaffected by $\varphi _0^{\rm A}$.

\emph{Proof:} Please refer to Appendix B.  \hfill {$\blacksquare$}

\begin{figure*}[ht] 
 	\centering
    \begin{align} \label{9}
    {R_{\rm{BD}}^{{\rm{A}}}} = \sum\limits_{m = 1}^M \frac{1}{M}\int\limits_{\mathbb{C} } {\frac{1}{{\pi \sigma _s^2}}\exp \left( { - \frac{{{{\left| {{y_{{\rm{MRC}}}} - g\frac{{m - 1}}{{M - 1}}\exp \left( {j{\varphi _0^A}} \right)} \right|}^2}}}{{\sigma _s^2}}} \right)} {{\log }_2}\left( {\frac{{\exp \left( { - \frac{{{{\left| {{y_{{\rm{MRC}}}} - g\frac{{m - 1}}{{M-1}}\exp \left( {j{\varphi _0^{\rm A}}} \right)} \right|}^2}}}{{\sigma _s^2}}} \right)}}{{\sum\limits_{i = 1}^M \frac{1}{M}\exp \left( { - \frac{{{{\left| {{y_{{\rm{MRC}}}} - g\frac{{i - 1}}{{M - 1}}\exp \left( {j{\varphi _0^{\rm A}}} \right)} \right|}^2}}}{{\sigma _s^2}}} \right)}}} \right) d{y_{{\rm{MRC}}}}
    \end{align}
     \hrulefill
\end{figure*}

\subsection{MPSK Modulation}
If the BD employs an MPSK modulation scheme, there are $M$ distinct  symbols, with the $m$-th symbol mapped to ${\Gamma _{m }}$, given by ${\Gamma _m} = {\alpha _0^{\rm P}}{\exp \left( {j{\varphi _m}} \right)}$, where ${\varphi _m} = {\varphi _0^{\rm P}} + \frac{{2\pi }}{M}\left( {m - 1} \right)$.
For simplicity in the analysis, we assume  $\varphi _{0}^{\rm P}\in  \left[ 0,\frac{2\pi}{M} \right)$ such that each symbol is distributed within the phase range of $\left[0,2\pi\right)$.
Then, using \eqref{3}, the PT's rate can be calculated as
\begin{align}\label{10}
{R_{\rm s}^{\rm{P}}} \!= \!\sum\limits_{m = 1}^M {\frac{1}{M}{{\log }_2}\left( {1 \!+\! \frac{{P\left|{h_1} \!+\! {h_2}{h_3}{\alpha _0^{\rm{P}}}{\exp \left( {j{\varphi _m}} \right)}\right|^2}}{{{\sigma ^2}}}} \right)}.
\end{align}

Here, we also   provide a Lemma to reveal the impact of MPSK modulation on the PT's rate.

\textbf{Lemma 4:} If BD employs the MPSK modulation scheme, the backscattering of BD may not lead to an enhancement in the PT's rate compared to the scenario without BD access. However, if the phases $\left\{ {{\varphi _m},m = 1,2,...,M} \right\}$ are carefully designed based on the  phases of $h_1$, $h_2$, and $h_3$, it can be ensured that the PT's rate can be enlarged by the access of BD.

\emph{Proof:} Please refer to Appendix C.  \hfill {$\blacksquare $}

{\emph{Remark 3.}} Lemma 4 reveals the following insight.
Unlike MASK, where the phase of the BD's symbol  within a  transmission block keeps unchanged, the phase ${\varphi _m}$ in MPSK follows a discrete uniform distribution ranging from 0 to 2$\pi$, which, however, does not guarantee an improvement in the PT's rate compared to the one without access of BD. This indicates that optimizing the phase ${\varphi _m}$ of MPSK is a key step in boosting the PT's rate.
Recall that the phases  $\left\{ {{\varphi _m},m = 1,2,...,M} \right\}$ of MPSK are determined by ${\varphi _0^{\rm P}}$, then we only need to optimize ${\varphi _0^{\rm P}}$ based on  $\theta_2 + \theta_3 - \theta_1$.

As mentioned above, the PT's rate is affected by the phases of MPSK. Since $M$ affects the phases of MPSK, in what follows,  we derive the PT's rate by assuming $M \to \infty $. Under this assumption,   the PT's rate can be written as
\begin{align}\label{11}
{R_{\rm s}^{\rm{P}}} &= {E_{\varphi} } \left[{\log _2}\left(1 + \frac{{P\left|{h_1} + {h_2}{h_3}{\alpha _0^{\rm{P}}} {e^{j\varphi }}\right|^2}}{{{\sigma ^2}}}\right)\right],
\end{align}
where $\varphi_m$ follows the  continuous  uniform distribution over the phase $\left[0, 2\pi\right)$.

Defining ${d_1} = 1 + \frac{{P\left( {|{h_1|^2} + |{h_2}{h_3}{\alpha _0^{\rm{P}}}|^2} \right)}}{{{\sigma ^2}}}$, $d_2 =\frac{{2P|{h_1}||{h_2}||{h_3}|{\alpha _0^{\rm{P}}}}}{{{\sigma ^2}}}$, and using  the  distribution of $\varphi$, \eqref{11} can rewritten as
\begin{align}\label{13}
{R_{\rm s}^{\rm{P}}} &= \frac{1}{{2\pi }}\int_0^{2\pi } {{{\log }_2}} \left(d_1 + d_2\cos \left({\theta_0} + \varphi \right)\right)d\varphi \nonumber\\
&  =  \frac{1}{{2\pi }}\int_0^{2\pi } {{{\log }_2}} \left(d_1 + d_2\cos u\right)du, \nonumber\\
&={\log _2}\left( {\frac{{{d_1} + \sqrt {{d_1}^2 - {d_2}^2} }}{2}} \right),
\end{align}
where the second equality holds from the variable substitution $u={{\theta_0 } + \varphi}$, and the third equality is derived  by using $d_1>d_2$, $\int_{0}^{\pi}\ln\left(l_1+l_2\cos x\right)dx = \pi\ln{\frac{l_1 +\sqrt{{l_1}^2-{l_2}^2}}{2}}, |l_1| > |l_2 | > 0$ \cite[eq. (4.224.9)]{zwillinger2007table} and $\int_0^\pi  {\ln } \left( {{l_1} + {l_2}\cos x} \right)dx = \int_\pi ^{2\pi } {\ln } \left( {{l_1} + {l_2}\cos x} \right)dx$.
The proof of $d_1>d_2$ is below: $ d_1 - d_2=1 + \frac{P}{\sigma^2} \left( |{h_1|}-|{h_2}{h_3}{\alpha _0^{\rm{P}}}|\right)^2>0$.

Based on \eqref{13}, we can obtain the following Lemma.

\textbf{Lemma 5:} If $M \to \infty $, then allowing  BD to backscatter information  ensures an increase in the PT's rate compared to that without BD access.

\emph{Proof:} Please refer to Appendix D.  \hfill {$\blacksquare $}

{\emph{Remark 4.}} Lemma 5 indicates that in MPSK, if the number of load impedances is sufficiently large, then the improvement in PT's rate  can always be  ensured for any  $\varphi_0^{\rm P}$ in the interval $[0,2\pi)$. Such an observation does not hold for MASK, as mentioned in Remark 2.
However, it is worth noting that the rate gain of PT can be further enlarged if $\varphi_0^{\rm P}$ can be carefully designed and adaptively changed based on  $\theta_2 + \theta_3 - \theta_1$.

\begin{figure*}[ht] 
 	\centering
    \begin{align} \label{14}
    {R_{\rm{BD}}^{{\rm{P}}}} = \sum\limits_{m = 1}^M \frac{1}{M}\int_ {\mathbb{C} } {\frac{1}{{\pi \sigma _s^2}}\exp \left( { - \frac{{{{\left| {{y_{{\rm{MRC}}}} - g \alpha _0^{\rm{P}} \exp \left( {j{\varphi _m}} \right)}
    \right|}^2}}}{{\sigma _s^2}}} \right)} {{\log }_2}\left( {\frac{{\exp \left( { - \frac{{{{\left| {{y_{{\rm{MRC}}}} - g \alpha _0^{\rm{P}} \exp \left( {j{\varphi _m}} \right)} \right|}^2}}}{{\sigma _s^2}}} \right)}}{{\sum\limits_{i = 1}^M \frac{1}{M}\exp \left( { - \frac{{{{\left| {{y_{{\rm{MRC}}}} - g \alpha _0^{\rm{P}} \exp \left( {j{\varphi _i}} \right)} \right|}^2}}}{{\sigma _s^2}}} \right)}}} \right) d{y_{{\rm{MRC}}}}
    \end{align}
    \hrulefill
\end{figure*}

Then, we derive the BD's rate under MPSK.
By substituting ${\Gamma _m} = {\alpha _0^{\rm{P}}}{\exp \left( {j{\varphi _m}} \right)}$ into \eqref{6a}, we obtain \eqref{14}, as shown at the top of the next page.

\textbf{Lemma 6:} When BD adopts the MPSK modulation, the rate of BD is not related to the phase $\varphi _0^{\rm P}$.

\emph{Proof:} Please refer to Appendix E.  \hfill {$\blacksquare $}

\section{Rate Maximization of the PT}
Section III highlights that the phase of MASK/MPSK has a significant impact on the PT's rate. This motivates us to  maximize the PT's rate by optimizing the phase of MASK/MPSK while ensuring the minimum transmission rate of the BD.

\subsection{The Design of MASK Modulation}
In this subsection, our goal is to maximize the rate of PT by optimizing the MASK phase.
The optimization problem is formulated as follows,
\begin{subequations}
    \begin{align}
\mathcal{P}_1: &\max_{\varphi _0^{\rm A}} R_{\rm s}^{\rm{A}}   \tag{\theparentequation} \label{eq:main} \\
\text { s.t. } &{R_{\rm{BD}}^{{\rm{A}}}}\geq R_{\min}^{\rm A} , \label{eq:con1}\\
&0 \leq \varphi _0^{\rm A} < 2\pi, \label{eq:con2}
    \end{align}
\end{subequations}
where $R_{\min}^{\rm A}$  represents the minimum transmission rate of BD under MASK modulation.

It appears that solving $\mathcal{P}_1$ is challenging due to the inclusion of $\varphi_0^{\rm A}$ in both $ R_{\rm s}^{\rm{A}} $ and $ R_{\rm{BD}}^{\rm{A}} $, whose expressions are complex. Fortunately,
using  Lemma 3, it is not hard to know the solution to $\mathcal{P}_1$ is equivalent  to that of the following optimization problem, given by
\begin{subequations}
    \begin{align}
    \mathcal{P}_{1.1}: &\max_{\varphi _0^{\rm A}} R_{\rm s}^{\rm{A}}   \tag{\theparentequation} \label{eq:main} \\
    \text { s.t. } &0 \leq \varphi _0^{\rm A} < 2\pi.
    \end{align}
\end{subequations}

Then, we only need to find a $\varphi_0^{\rm A}$ within the interval $[0, 2\pi)$ that maximizes ${R_{\rm{s}}^{{\rm{A}}}}$.
Substituting $\theta_0=\theta_2 + \theta_3 - \theta_1$ into  \eqref{8} and after some mathematical operations, we rewrite ${R_{\rm{s}}^{{\rm{A}}}}$ as
\begin{align} \notag\label{17}
{R_{\rm{s}}^{{\rm{A}}}}&=\sum\limits_{m = 1}^M \!\!\frac{1}{M}{{\log }_2}\!\!\left(\!\! 1 \!\!+ \!\!\frac{{P\!\!\left( \!\!{{{\left| {{h_1}} \right|}^2}\!\! +\!\! {{\left| {{h_2}{h_3}\frac{{m - 1}}{{M - 1}}} \right|}^2}} \right)}}{{{\sigma ^2}}} \right.\\
&\left.+ \frac{{2P\left| {{h_1}} \right|\left| {{h_2}} \right|\left| {{h_3}} \right|\frac{{m - 1}}{{M - 1}}\cos ({\theta _0}{\rm{ + }}{\varphi _0^{\rm A}})}}{{{\sigma ^2}}} \right).
\end{align}

{\textbf{Lemma 7:} The  optimal $\varphi _0^{\rm A^*}$ to $\mathcal{P}_{1.1}$ is selected from the set
\begin{align} \label{20}
\left\{ {\varphi _0^{\rm A} \left| {{\varphi _0^{\rm A}} = 2\lambda \pi - {\theta _0}, \lambda\in \mathbb{Z}, 0 \le {\varphi _0^{\rm A}} < 2\pi } \right.} \right\}.
\end{align}
where $\mathbb{Z}$ denoting the set of integers.

{\emph{Proof:} According to \eqref{17}, $R_{\rm s}^{\rm A}$ is maximized when ${\cos ({\theta _0} + {\varphi _0^{\rm A}})}=1$ must hold.
This implies that ${\theta _0}+ {\varphi _0^{\rm A}} = 2\lambda \pi $.
Combining this with the constraint ${\varphi _0^{\rm A}} \in [0,2\pi )$, {\color{black} the optimal phase $\varphi _0^{\rm A}{^*}$ for $\mathcal{P}_{1.1}$ is obtained.}

Substituting the optimal $\varphi_0^{\rm{A}^*}$ into \eqref{17}, we obtain the maximum PT's rate when the MASK is used at BD, given by
\begin{align}
{R_{\rm{s}}^{{\rm{A}}}}&\!=\!\sum\limits_{m = 1}^M \frac{1}{M}{{\log }_2}\!\left(\! 1 \!+ \!\frac{{P\!\left( \!{{{\left| {{h_1}} \right|}}\! +\! {{\left| {{h_2}{h_3}\frac{{m - 1}}{{M - 1}}} \right|}}} \right)^2}}{{{\sigma ^2}}} \right).
\end{align}


\subsection{The Design of MPSK Modulation}
In this subsection, we aim to maximize the PT's rate by optimizing the phase $\varphi _0^{\rm P}$. The corresponding optimization problem is formulated as
%
%
\begin{subequations}
    \begin{align}
\mathcal{P}_{2}: &\max_{\varphi _0^{\rm P}} R_{\rm s}^{\rm{P}}   \tag{\theparentequation} \label{eq:main} \\
\text { s.t. } &{R_{\rm{BD}}^{{\rm{P}}}}\geq R_{\min}^{\rm{P}}, \label{eq:con1}\\
&0 \leq \varphi _0^{\rm P} < \frac{2\pi}{M}, \label{eq:con2.1}
    \end{align}
\end{subequations}
where $R_{\min}^{\rm P}$ represents the minimum transmission rate of BD under MPSK modulation.

Since  ${\varphi _0^{\rm P}}$ is included in  $R_{\rm s}^{\rm{P}}$ and ${R_{\rm{BD}}^{{\rm{P}}}}$, whose expressions are complex, it is challenging to solve $\mathcal{P}_{2}$.
Lemma 6 indicates that  $\mathcal{P}_{2}$  can be equivalently transformed into  the following problem, given by
\begin{subequations}
    \begin{align}
\mathcal{P}_{2.1}: &\max_{\varphi _0^{\rm P}} R_{\rm s}^{\rm{P}}   \tag{\theparentequation} \label{eq:main23.1} \\
\text { s.t. } &0 \leq \varphi _0^{\rm P} < \frac{2\pi}{M}. \label{eq:con23.2}
    \end{align}
\end{subequations}

To find an optimal ${\varphi _0^{\rm P}}$ that maximizes $R_{\rm s}^{\rm{P}} $, we rewrite  \eqref{10} as
\begin{align}\label{21}
&R_{\rm s}^{\rm{P}} = \sum\limits_{m = 1}^M \frac{1}{M}{{\log }_2}\left( 1 + \frac{{P\left( {{{\left| {{h_1}} \right|}^2} + {{\left| {{h_2}{h_3}} \right|}^2}{\alpha _0^{\rm{P}}}^2} \right)}}{{{\sigma ^2}}} \right. \nonumber\\
&\left.+ \frac{{2P\left| {{h_1}} \right|\left| {{h_2}} \right|\left| {{h_3}} \right|{\alpha _0^{\rm{P}}} \cos \left( {{\theta _0} + {\varphi _0^{\rm P}} + \frac{{2\pi }}{M}\left( {m - 1} \right)} \right)}}{{{\sigma ^2}}} \right).
\end{align}

It can be seen from \eqref{21} that $R_{\rm s}^{\rm{P}}$ is a  sum-log-cos function and thus non-convex.
Although successively convex approximation can be used to approximate \eqref{21} to a linear one and find a locally optimal $\varphi_0^{\rm P}$, such an approach does not guarantee  to obtain the globally optimal solution.
To address this issue, in what follows, we apply mathematical induction to find the optimal $\varphi _0^{{\rm P}^*}$ that maximizes  $R_{\rm s}^{\rm{P}}$, and the  result is summarized below.

{\textbf{Lemma 8:} The  optimal $\varphi _0^{\rm P^*}$ to $\mathcal{P}_{2.1}$ is selected from the  set
\begin{align} \label{24}
\left\{{{\varphi _0^{\rm P}}\left| {\frac{\pi }{M} + \frac{{2\eta  \pi }}{M}- {\theta _0},\eta \in \mathbb{Z}, 0 \le  {\varphi _0^{\rm P}}<\frac{{2\pi }}{M}} \!\right.} \right\}.
\end{align}

{\emph{Proof:} The phase $\varphi _0^{\rm P}$ that maximizes $R_{\rm s}^{\rm{P}}$ is given by ${\varphi _0^{\rm P}} = \frac{\pi }{M} + \frac{{2\eta \pi }}{M} - {\theta _0}$, and the detailed proof can be found in  Appendix F.
Combining it with  constraint \eqref{eq:con23.2}, we obtain the optimal phase $\varphi _0^{\rm P^*}$ for $\mathcal{P}_{2.1}$. \hfill {$\blacksquare $}.

Substituting  $\varphi_0^{{\rm P}^*}$ into \eqref{21}, we obtain the maximum PT's rate when the BD adopts MPSK as follows,
\begin{align}
&R_{\rm{s}}^{{\rm{P}}}{\rm{ }} = \sum\limits_{m = 1}^M {\frac{1}{M}} {\log _2}\left( {1 + \frac{{P\left( {{{\left| {{h_1}} \right|}^2} + {{\left| {{h_2}{h_3}} \right|}^2}{\alpha _0^{\rm{P}}}^2} \right)}}{{{\sigma ^2}}}} \right.\nonumber\\
&\left. { \!+\! \frac{{2P\left| {{h_1}} \right|\left| {{h_2}} \right|\left| {{h_3}} \right|{\alpha _0^{\rm{P}}}\cos \left( {{\theta _0} + \varphi _0^{{\rm P}^*}  + \frac{{2\pi }}{M}\left( {m - 1} \right)} \right)}}{{{\sigma ^2}}}} \right).
\end{align}

{\emph{Remark 5.}}  So far, we have derived the optimal phase of the MASK and MPSK for SBC, which can improve the PT's rate when the BD backscatters information. It is evident that the design of MASK/MPSK in SBC differs significantly from conventional communications, where  the phase of MASK/MPSK does not impact transmission performance. However, in SBC, because the  signal reflected by  BD contains both the BD and PT symbols, the phase of MASK/MPSK at the BD affects $ h_{{\rm{eq}},m}$, which is a variable influencing the PT's rate. Furthermore, we observe that the optimal phase of MASK is solely determined by $\theta_2+\theta_3-\theta_1$, whereas for MPSK, the optimal phase depends on both $\theta_2+\theta_3-\theta_1$  and $ M $.

{\color{black}\subsection{Practical Implementation of Near-Optimal Modulation}
Here we address their practical implementation in BDs. A key challenge arises because practical BDs can only realize a finite set of discrete load impedances, and thus discrete reflection phases, whereas the theoretical optimum may require a continuous phase value. This subsection elaborates on a practical circuit design that uses discrete phases to closely approximate the optimal performance.

For the MASK modulation, we need to pre-establish $N \times M$ load impedances to  offer $N$ phases under a given amplitude of the complex reflection coefficient.
Specifically, we pre-design $N$ sets of load impedances.
The $i$-th set corresponds to a phase $\varphi^{i,{\rm A}} = \frac{{2\pi }}{N}\left( {i - 1} \right)$, where $i = 1,2, \ldots N $.
Each set contains contains $M$ impedances, which correspond to the $M$ distinct amplitude levels in MASK modulation.
The reflection coefficient for the $m$-th impedance within the $i$-th set is ${\Gamma _{m}} = \frac{{m  - 1}}{{M - 1}}\exp \left(j{\varphi^{i,{\rm A}}}\right)$.
This design ensures that each phase group is a self-contained MASK modulation.  {\emph{ At the beginning of each transmission block,  the backscatter device (BD) first determines the optimal phase $\varphi _0^{\rm A^*}$ based on the phases of channel coefficients $h_1$, $h_2$ and $h_3$, and  selects the impedance set $i$ whose phase $\varphi^{i,{\rm A}}$ is closest to $\varphi _0^{\rm A^*}$.
Then, within the chosen set, it directly switches to the specific impedance corresponding to the intended amplitude for backscattering information to the receiver.}}
This process ensures that the backscattered signal operates with both the correct amplitude and a phase close to the theoretical optimum.
It is worth noting that when the amplitude of MASK is zero (i.e., $m=1$), no reflection occurs regardless of the phase value.
This property allows us to further reduce the total number of required load impedances.
Specifically, instead of implementing $N \times M$ impedances, we can optimize the design to require only $N \times \left( M - 1 \right) + 1$ impedances.
This reduction is achieved by sharing a single common impedance for the zero-amplitude state across all phase sets, as illustrated in Fig. \ref{fig21}.

\begin{figure}
\centering
  \includegraphics[width=0.4\textwidth]{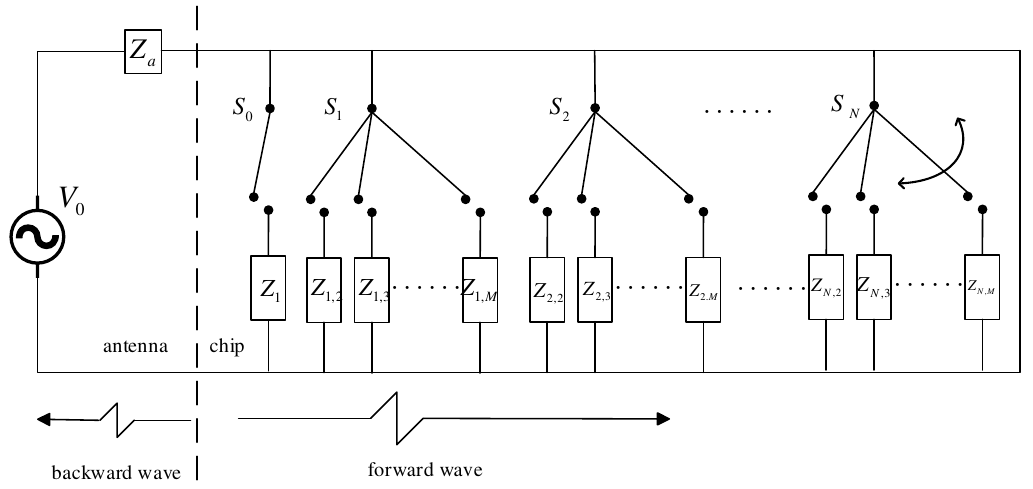}\\
  \caption{MASK circuit design.}\label{fig21}
\end{figure}

\begin{figure}
\centering
  \includegraphics[width=0.4\textwidth]{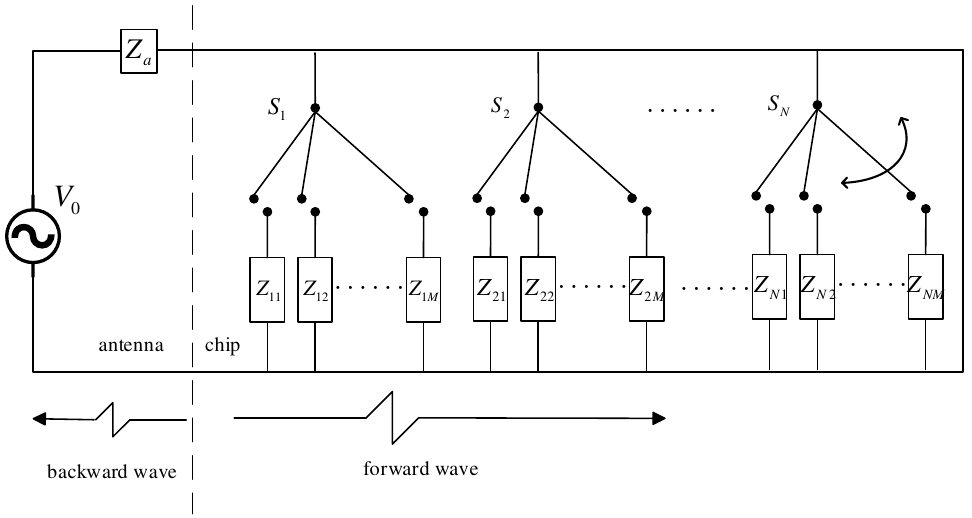}\\
  \caption{MPSK circuit design.}\label{fig22}
\end{figure}

\begin{figure*}
	\centering
	\begin{minipage}[t]{0.3\linewidth}    
		\centering    
		\includegraphics[width=1\linewidth]{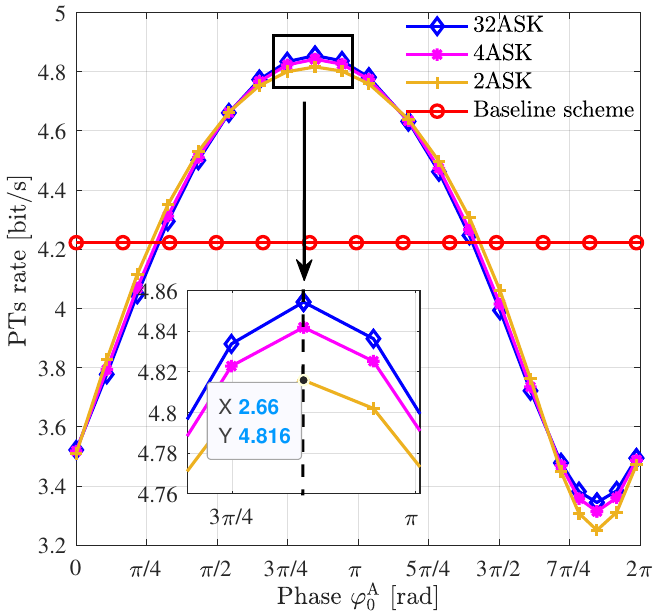}
		\caption{{\color{black}PT's rate versus phase $\varphi_0^{\rm A}$: ASK  modulation, with $\ell _1$, $\ell _2$, $\ell _3$.}}   
		\label{fig3}
	\end{minipage}%
        \hspace{1mm}    
	\begin{minipage}[t]{0.31\linewidth}
		\centering
		\includegraphics[width=1\linewidth]{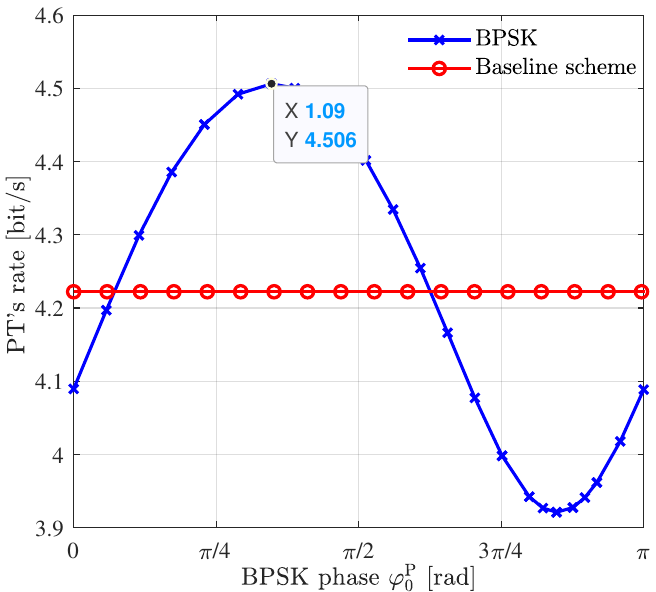}
		\caption{{\color{black}PT's rate versus phase $\varphi_0^{\rm P}$: BPSK modulation, with $\ell _1$, $\ell _2$, $\ell _3$.}}
        \label{fig4}
	\end{minipage}
        \hspace{1mm}
	\begin{minipage}[t]{0.31\linewidth}
		\centering
		\includegraphics[width=1\linewidth]{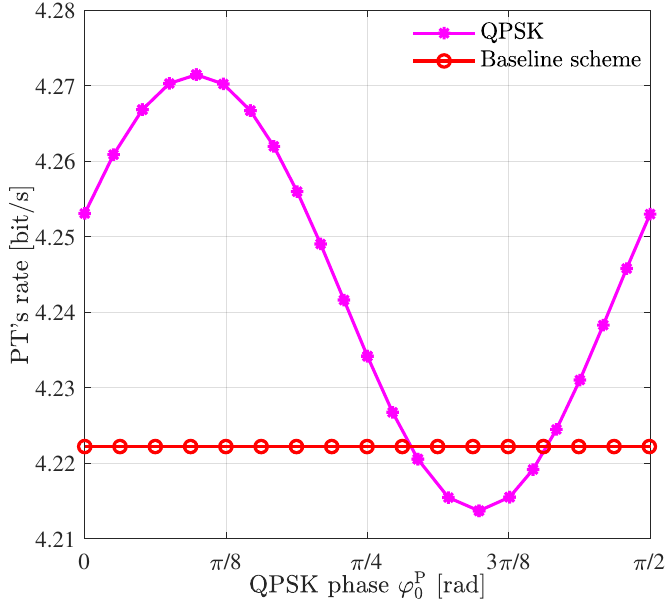}
		\caption{{\color{black}PT's rate versus phase $\varphi_0^{\rm P}$: QPSK modulation, with $\ell _1$, $\ell _2$, $\ell _3$.}}
        \label{fig5}
	\end{minipage}
\end{figure*}

For MPSK modulation, the objective is to apply a near-optimal initial phase $\varphi _0^{\rm P^*}$.
Similarly, we pre-design $N$ sets of load impedances.
The $i$-th set corresponds to an initial phase $\varphi^{i,{\rm P}} = \frac{{2\pi \times \left( {i - 1} \right)}}{M \times N}$, and each set contains $M$ impedances realizing the MPSK symbol phases relative to ${\varphi^{i,{\rm P}}}$, where $i = 1,2, \ldots N $.
The reflection coefficient for the $m$-th impedance within the $i$-th set is ${\Gamma _m} = {\alpha _0^{\rm P}}{\exp \left( {j \left({{\varphi^{i,{\rm P}}} + \frac{{2\pi }}{M}\left( {m - 1} \right)}\right) }\right)}$, where $\alpha _0^{\rm P}$ denotes the amplitude of MPSK.
{\emph{At the beginning of each transmission block, the BD selects the impedance set whose initial phase $\varphi^{i,{\rm P}}$ most closely approximates $\varphi _0^{\rm P^*}$. Then,  the MPSK modulation is performed by switching among the $M$ impedances within the selected set for backscattering information to the receiver.}}
The circuit structure enabling this functionality, based on load impedance modulation, is illustrated in Fig. \ref{fig22}.

We acknowledge that this design requires $N \times M$ distinct impedances and this increases the hardware cost.
However, we argue that this increase is highly manageable and justified for two key reasons.
First, the required  impedances are passive and remain extremely low-cost. The fundamental architecture of the BD is preserved, with complexity confined to the switching logic.
Second, as discussed in the next paragraph (see Fig. \ref{fig23} and Fig. \ref{fig24}), the PT's rate  for even a small value of $N$ is very close to that of the infinite-$N$ case, indicating diminishing returns for further hardware investment.}

\section{Simulation Results}
In this section, simulation results are presented to evaluate the influence of the BD's modulation scheme on the PT's rate.
The simulation parameters are set as follows unless otherwise specified.
The channel coefficients are defined as $h_i=\sqrt{\mu _i}\ell _i$, where $i = 1, 2, 3$.
Here $\mu_i$ represent the large-scale path losses, and $\ell_i$  denote the small-scale fading components\footnote{Here, the small-scale fading components involved in the simulation are set as follows: $\ell _1 = 0.3421 - 0.4988i$, $\ell _2 = -0.0139 - 0.4378i$ and $\ell _3 =-0.5246 - 1.0546i$, $\ell _1^{\prime} = 0.2651 + 0.0031i$, $\ell _2^{\prime} = -1.2621 + 0.0425i$, $\ell _3^{\prime} =   -0.3110 - 0.7787i$. Notably, the results remain robust even as these parameters vary randomly.} that follow  $\ell \sim \mathcal{CN} \left(0,1\right)$.
The distances between the PT-R, PT-BD, BD-R are set as $d_1 = d_2 = 200$ m, and $d_3 = 0.36$ m.
The path losses are calculated using the model proposed in \cite{8907447}, which is expressed as
\begin{align}
\mu _1=\frac{\lambda _{c}^{2}G_p G_r }{\left ( 4\pi  \right ) ^{2}d_1^{v_1} } , \mu _2=\frac{\lambda _{c}^{2}G_pG_b }{\left ( 4\pi  \right ) ^{2}d_2^{v_2} }, \mu _3=\frac{\lambda _{c}^{2}G_b G_r }{\left ( 4\pi  \right ) ^{2}d_3^{v_3} },\nonumber
\end{align}
where $\lambda _{c}=0.33$ m is the wavelength of the carrier signal, corresponding to 900 MHz, the path loss exponent is $v=3.5$, and the antenna gains are $G_p = G_r = G_b=6$ dB.
Therefore, the calculated path losses are $\mu _1=\mu _2=-100$ dB and $\mu _3=-4$ dB.
Additionally, we assume the PT's transmit power is $P=0.05$ W, the noise power at the PR is $-100$ dBm, and $\alpha_0^{\rm P}=0.9$.
Beyond these parameters, we consider the case without BD access as a baseline scheme for comparison.





{\color{black}Fig. \ref{fig3} depicts the PT's rate $R_s$ versus the MASK phase $\varphi _0^{\rm A}$.
The optimal phase derived from theoretical analysis is consistent with the phase that maximizes the PT's rate in simulations.
Moreover, when the phase is incorrectly chosen, the PT's rate is lower than that without BD access,
which confirms the importance of phase selection.
Besides, it is observed that when BD employs ASK modulation, the optimal phase remains constant regardless of the modulation order $M$.
This indicates that for the MASK modulation, the optimal phase for maximizing the PT's rate is uniquely determined by the channel phase, which is in agreement with the Remark 1.
As the phase approaches the optimal value, the PT's rate increases, while it decreases when the phase deviates from the optimal value in $\left [ 0,2\pi  \right)$.}

{\color{black}Figs. \ref{fig4} and \ref{fig5} compares the PT's rate versus the phase $\varphi _0^{\rm P}$ under BPSK and QPSK modulations.
As shown in Figs. \ref{fig4} and \ref{fig5}, the optimal phase predicted by theoretical analysis
aligns with the phase that maximizes the PT's rate in simulations, validating the correctness of \eqref{24}.
Additionally, the results reveal that when the phase is not correctly selected, the PT's rate $R_{\rm s}^{\rm P}$ is lower compared to the case without BD access.
This underscores the importance of phase optimization with MPSK modulation in SBC systems.
A comparing between Figs. \ref{fig4} and \ref{fig5} further reveals that the optimal phase $\varphi_0^{\rm P}$ in PSK modulation varies with different modulation orders $M$.
This observation that contrasts with MASK modulation, where the optimal phase $\varphi_0^{\rm A}$ remains constant.
The difference stems from the fact that in MPAK modulation, information is mapped to varying phases, whereas in MASK, the phase is kept fixed.}


\begin{figure*}
	\centering
	\begin{minipage}[t]{0.3\linewidth}    
		\centering    
		\includegraphics[width=1\linewidth]{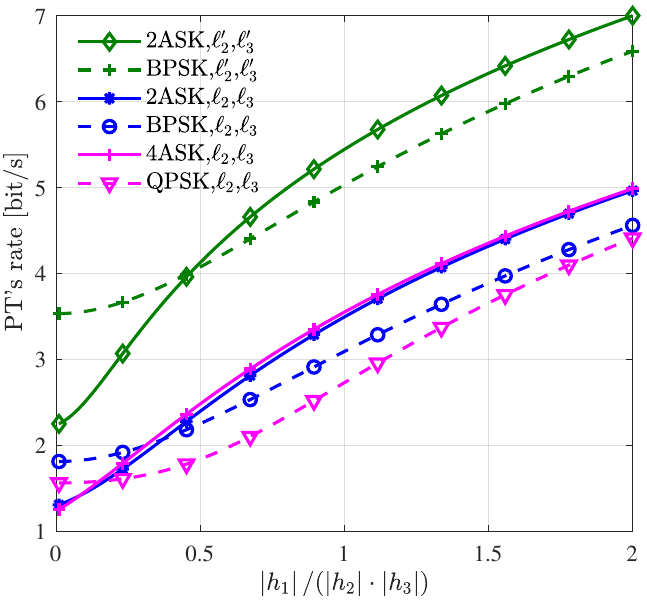}
		\caption{{\color{black}PT's rate versus phase $\varphi_0^{\rm P}$: QPSK modulation, with $\ell _1$, $\ell _2$, $\ell _3$.}}   
		\label{fig8}
	\end{minipage}%
        \hspace{1mm}    
	\begin{minipage}[t]{0.31\linewidth}
		\centering
		\includegraphics[width=1\linewidth]{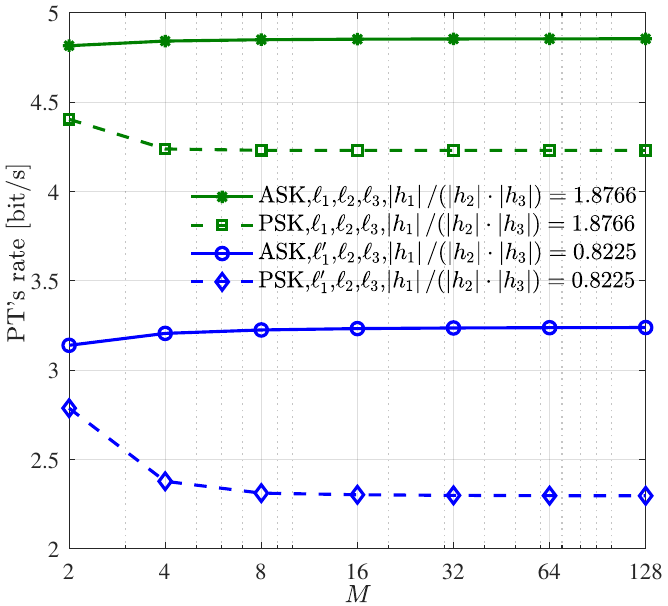}
		\caption{PT's rate versus $M$ under the optimal $\varphi _0^{\rm A}$ and $\varphi _0^{\rm P}$.}
        \label{fig9}
	\end{minipage}
        \hspace{1mm}
	\begin{minipage}[t]{0.31\linewidth}
		\centering
		\includegraphics[width=1\linewidth]{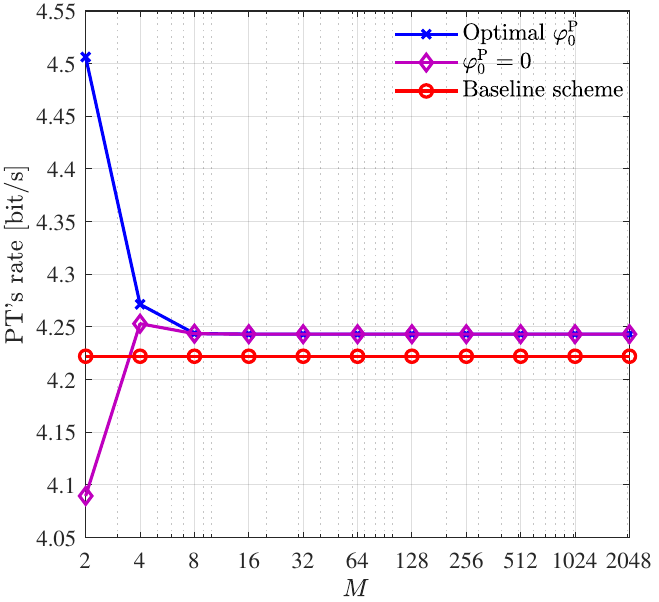}
		\caption{PT's rate versus $M$ under the PSK optimal $\varphi _0^{\rm P}$, with $\ell _1$, $\ell _2$, $\ell _3$.}
        \label{fig10}
	\end{minipage}
\end{figure*}

To investigate which modulation scheme more effectively enhances the PT's rate, we compare the performance of ASK and PSK at their optimal phases $\varphi_0^{\rm A}$ and $\varphi_0^{\rm P}$, respectively. The results are presented in Figs. \ref{fig8} and \ref{fig9}.
In Fig. \ref{fig8}, the primary variable of interest is the channel ratiochannel ratio $\frac{\left| h_1 \right|}{ \left| h_2 \right| \cdot \left| h_3 \right|}$.
The amplitude of $h_1$  is not explicitly specified, as it is determined by the predefined amplitudes of $h_2$, $h_3$, and the desired ratio.
Conversely, Fig. \ref{fig9} investigates the impact of varying modulation orders $M$ on the PT's rate.
In addition, to facilitate comparison, we specify that ASK and PSK modulations have equal average transmit power for the same modulation order $M$ in Figs. \ref{fig8} and \ref{fig9}.


Fig. \ref{fig8} illustrates the effect of the channel ratio $\frac{\left| h_1 \right|}{ \left| h_2 \right| \cdot \left| h_3 \right|}$ on the PT's rate for both MASK and MPSK modulations with optimal phase.
For a given $M$, there exists a unique channel ratio $r_0$ at which the PT's rates under ASK and PSK modulations are equal.
Specifically, when the ratio is below $r_0$, PSK modulation achieves a higher PT's rate than ASK; conversely, ASK outperforms PSK when the ratio exceeds $r_0$.
This indicates that when the direct link $h_1$ is weak, the PT's rate of ASK may be lower than that of PSK in SBC.
Additionally, for different values of $M$, the PT's rates under ASK and PSK modulations are equal at distinct channel ratio values.
As $M$ increases, these channel ratio becomes smaller.


{\color{black}Fig. \ref{fig9} plots the relationship between the PT's rate and the modulation orders $M$ of ASK and PSK, where the phase in each channel parameter is set to the optimal value.
By observing the PT's rates under PSK and ASK modulation for the two sets of channel conditions,
the results are consistent with the conclusions derived from Fig. \ref{fig8}.
Specifically, when the channel ratio is relatively high, ASK modulation outperforms PSK in enhancing the PT's rate.
Additionally, it can be observed that for ASK modulation, the PT's rate increases as $M$ grows, whereas for PSK modulation, the rate decreases as $M$ increases.}

Fig. \ref{fig10} shows the trend of the PT's rate as $M$ varies under the optimal phase for MPSK.
Under the optimal phase $\varphi_0^{\rm P}$, the PT's rate is significantly higher than that in the case without BD access.
By comparing the PT's rates with the optimal phase and not optimal phases, it can be observed that as the $M$ increases, the PT's rate gradually converge and eventually coincide.
This is because, as $M$ increases, the influence of the phase $\varphi_0^{\rm P}$ on the PT's rate under PSK modulation diminishes.
Furthermore, when $M$ is sufficiently large, the PT's rate stabilizes.
Thus, it is reasonable to infer that as $M \to \infty$, the PT's rate remains higher than that of the baseline scheme, which is consistent with the theoretical results derived from Lemma 5.

\begin{figure}
  \centering 
  \includegraphics[width=0.35\textwidth]{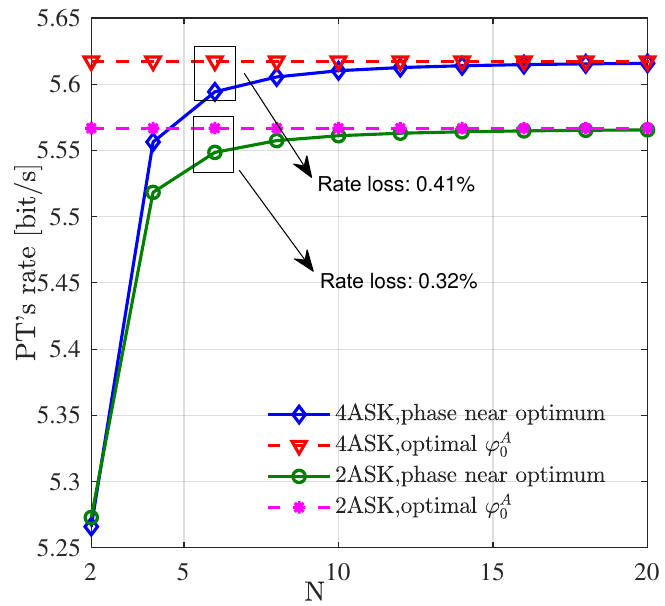}\\  
  \captionsetup{justification=centering} 
  \caption{{\color{black}MASK Modulation: PT's rate vs. $N$.}}\label{fig23}
\end{figure}

\begin{figure}
  \centering 
  \includegraphics[width=0.35\textwidth]{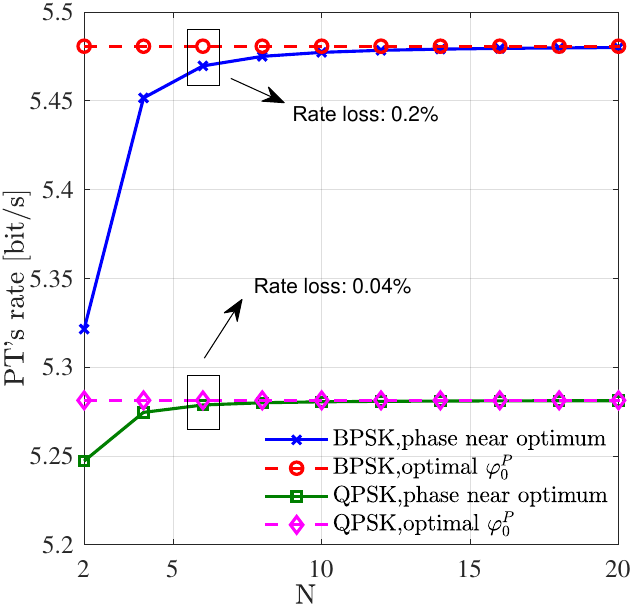}\\  
  \captionsetup{justification=centering} 
  \caption{{\color{black}MPSK Modulation: PT's rate vs. $N$}}\label{fig24}
\end{figure}

{\color{black}Figs. \ref{fig23} and \ref{fig24} depict the variation of the average rate of the PT with the number of phase partitions under near-optimal and optimal phases across multiple channels for MASK and MPSK modulations, respectively.
It can be observed that as $N$ increases, the rate of the near-optimal phase scheme rapidly converges to the optimal rate.
For both modulation schemes, the rate loss becomes extremely small when $N \ge 6$, and the rapid convergence to the optimal performance with a relatively small $N$ (i.e., $N \ge 8$) demonstrates both the theoretical soundness and practical feasibility of our proposed approach.
These findings thus offer valuable design guidelines for implementing efficient, low-complexity symbiotic backscatter communication systems.}

\begin{figure}
  \centering 
  \includegraphics[width=0.35\textwidth]{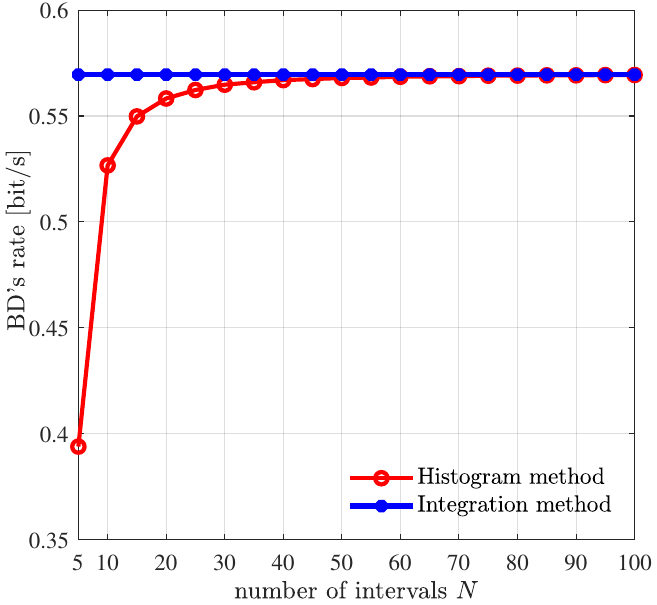}\\  
  \captionsetup{justification=centering} 
  \caption{{\color{black}BD's rate versus number of intervals $N$ under ASK modulation.}}\label{fig41}
\end{figure}

\begin{figure}
  \centering 
  \includegraphics[width=0.35\textwidth]{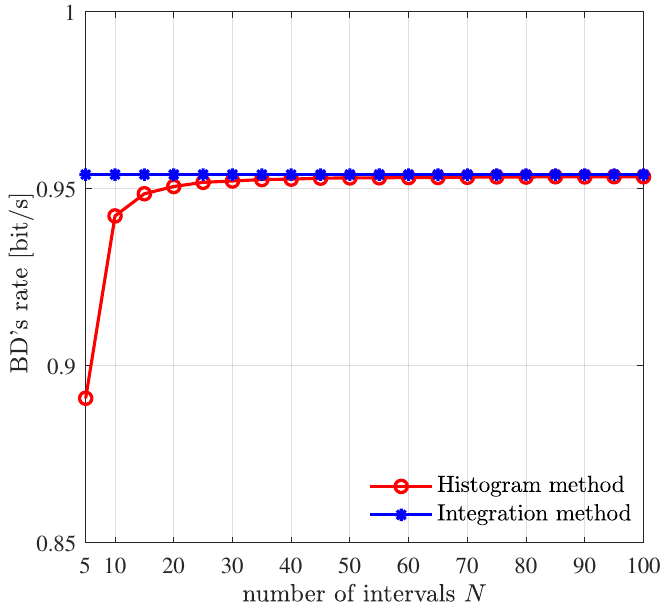}\\  
  \captionsetup{justification=centering} 
  \caption{{\color{black}BD's rate versus number of intervals $N$ under BPSK modulation.}}\label{fig42}
\end{figure}

Figs. \ref{fig41} and \ref{fig42} illustrate the variation of the BD's rate with the number of intervals
$N$ under ASK modulation and BPSK modulation, respectively. It can be observed that as the number of intervals $N$ increases, the BD's rate gradually increases.
When $N$ grows to a certain extent, the results of the histogram method tend to stabilize and align perfectly with those of the integration method.
This not only validates the accuracy of the histogram method in approximating complex plane integrals  but also indicates that the histogram method can achieve a high-precision approximation of the integration method with only a small number of intervals.

\section{Conclusions}
In this paper, we have studied the impact of the BD adopting MASK or MPSK modulation on SBC and  derived expressions for the PT's rate and BD's rate when the BD adopts MASK and MPSK, respectively.
We have shown that the phase of MASK/MPSK significantly impacts the PT's rate, while it remains independent of the BD's rate, as summarized in Remarks 1-4. Additionally, {\color{black}we have developed optimal phase designs for MASK and MPSK modulation to maximize the PT's rate, supported by a practical circuit that enables BDs to realize near-optimal performance using a finite set of discrete load impedances.}
Simulation results have been provided to validate the theoretical findings, confirming the following key points. First, when the direct link is not significantly weaker than the backscatter link, ASK outperforms PSK in terms of PT's rate, and vice versa. Second,  under the MASK scheme, the PT's rate increases with $M$; while under the MPSK scheme, the PT's rate decreases as $M$ increases. Third, carefully designing the  phase of MPSK can improve the PT's rate, however, as $M \to \infty $, the advantage introduced by the optimal phase becomes negligible. Based on the above findings and considering the  low-complexity BD, we have also confirmed that the low-order modulation AKS is a better choice for the BD since it provides a higher rate for the PT.

\section*{Appendix A}
The rate gain of PT can be expressed as
\begin{align}\label{b1}
\Delta {R_{\rm s}^{\rm{A}}}=\int_{0}^{1} {{{\log }_2}\left( {1 + \frac{{P|{h_1} + {h_2}{h_3}\alpha \exp \left( {j{\varphi _0^{\rm A}}} \right){|^2}}}{{{\sigma ^2}}}} \right)\!}d\alpha - R_p. \tag{A.1}
\end{align}

If $\theta_2 + \theta_3-\theta_1+{\varphi _0^{\rm A}} = \pm\pi $, and $\left| {{h_1}} \right|>\left| {{h_2}} \right|\left| {{h_3}} \right|$, \eqref{b1} can be rewritten as
\begin{align}\label{b2}\notag
\Delta {R_{\rm s}^{\rm{A}}}&=\int_{0}^{1} {{\log }_2}\left ( 1 + \frac{P \left ( \left | h_1 \right | - \alpha\left | h_2h_3 \right |   \right )^2 }{\sigma ^{2} }  \right )d\alpha  - R_p\\
 &< \int_0^1 {{{\log }_2}} \left( {1 + \frac{{P{{\left| {{h_1}} \right|}^2}}}{{{\sigma ^2}}}} \right)d\alpha  - {R_p} = 0. \tag{A.2}
\end{align}

On the other hand, when $\theta_2 + \theta_3 - \theta_1 + {\varphi _0^{\rm A}} = 0 $, we have
\begin{align}\notag
\Delta {R_{\rm s}^{\rm{A}}}&=\int_{0}^{1} {{\log }_2}\!\left ( 1 + \frac{P \left ( \left | h_1 \right | +\alpha\left | h_2h_3 \right |   \right )^2 }{\sigma ^{2} }  \right )d\alpha  -R_p\\
&> \int_0^1 {{{\log }_2}} \left( {1 + \frac{{P{{\left| {{h_1}} \right|}^2}}}{{{\sigma ^2}}}} \right)d\alpha  - {R_p} = 0. \tag{A.3}
\end{align}

The proof is complete.
\section*{Appendix B}
According to \eqref{9}, the received signal $y_{\rm{MRC}}$ in polar coordinates is expressed as $y_{\rm{MRC}}=r{e^{j\phi }}$.
Therefore, \eqref{e1} can be obtained, as shown at the top of the next page.

\begin{figure*}[ht] 
 	\centering
    \begin{align} \label{e1}
    R_{\rm{BD}}^{\rm{A}} &\!= \!\frac{1}{{M\!\pi\! \sigma _s^2}}\!\!\sum\limits_{m = 1}^M \!\! {\int\limits_0^{2\pi } {\int\limits_0^{ + \infty } {{\!\!\exp\!\!\left({ \!\!- \!\frac{{{r^2} \!+\! {{\!\left( \!{g\frac{{m \!-\! 1}}{{M\! -\! 1}}} \!\!\right)}^2} \!- \!2rg \frac{{m\! -\! 1}}{{M \!- \!1}}\cos \!\left(\! {\phi \! - \!{\varphi _0^{\rm A}}} \!\right)\!\!}}{{\sigma _s^2}}}\right)}} } }{\log _2}\!\!\left( \!\!{\frac{{{\exp\left({ - \frac{{\left(\! {g\frac{{m - 1}}{{M - 1}}} \!\!\right)^2 - 2rg \frac{{m - 1}}{{M - 1}}\cos\! \left(\! {\phi \! - \!{\varphi _0^{\rm A}}} \!\right)\!}}{{\sigma _s^2}}}\right)}}}{{\frac{1}{M}\!\sum\limits_{i = 1}^M {{\exp\!\left({ \!- \!\frac{{{{\left( {g\frac{{i - 1}}{{M - 1}}} \right)}^2} \!-\! 2rg\frac{{ i - 1}}{{M - 1}}\cos \!\left( \!{\phi \! - \!{\varphi _0^{\rm A}}} \!\right)\!}}{{\sigma _s^2}}}\!\right)\!}} }}} \!\!\right)\!\!rdrd\phi \tag{B.1}
    \end{align}
    \hrulefill
\end{figure*}

By introducing the variable ${\phi ^\prime } = \phi  - {\varphi _0^{\rm A}}$, and  noting that $\phi  - {\varphi _0^{\rm A}}$ only appears in $\cos(x)$, \eqref{e1} can be further simplified to \eqref{e2}, as shown at the top of the next page. {\color{black}Although the substitution ${\phi ^\prime } = \phi  - {\varphi _0^{\rm A}}$ shifts the limits of integration from $\left[ {0,2\pi } \right]$ to $\left[ {-\varphi_0^A,2\pi - \varphi_0^A } \right]$, the integrand is periodic in $\phi^{\prime}$ with period $2\pi$ due to the presence of $\cos {\phi ^\prime}$. As a result, the integral over any complete period of length $2\pi$ is equivalent. Therefore, the limits in \eqref{e2} can be validly taken as 0 to $2\pi$ without affecting the result.}

\begin{figure*}[ht] 
 	\centering
    \begin{align}\notag\label{e2}
    R_{\rm{BD}}^{\rm{A}} &\!\!= \!\!\frac{1}{{M\!\pi\! \sigma _s^2}}\!\!\sum\limits_{m = 1}^M \!\! {\int\limits_0^{2\pi } {\int\limits_0^{ + \infty } {{\!\!\exp\!\!\left({ \!\!- \!\frac{{{r^2} \!+\! {{\!\left( \!{g\frac{{m \!-\! 1}}{{M\! -\! 1}}} \!\!\right)}^2} \!- \!2rg \frac{{m\! -\! 1}}{{M \!- \!1}}\cos \!{\phi ^\prime}\!}}{{\sigma _s^2}}}\right)}} } }{\log _2}\!\!\left( \!\!{\frac{{{\exp\!\left({\! -\! \frac{{\left(\! {g\frac{{m - 1}}{{M - 1}}} \!\!\right)^2 - 2rg \frac{{m - 1}}{{M - 1}}\cos\! {\phi ^\prime} \!}}{{\sigma _s^2}}}\right)}}}{{\frac{1}{M}\!\sum\limits_{i = 1}^M {{\exp\!\left({ \!- \!\frac{{{{\left( {g\frac{{i - 1}}{{M - 1}}} \right)}^2} \!-\! 2rg\frac{{ i - 1}}{{M - 1}}\cos \! {\phi ^\prime} \!}}{{\sigma _s^2}}}\!\right)\!}} }}} \!\!\right)\!\!rdrd\phi. \tag{B.2}
    \end{align}
    \hrulefill
\end{figure*}

It can be seen from \eqref{e2}  that ${R_{\rm{BD}}^{\rm{A}}}$ is independent of the phase ${\varphi _0^{\rm A}}$ of $\Gamma_m$, indicating that the BD rate with MASK modulation is unaffected by the symbol phase.
\section*{Appendix C}
This Appendix provides two examples to illustrate that  the access of the BD degrades the PT's rate and that the access of the BD improves the PT's rate, respectively.

\subsubsection{Example 1} In this example, we assume that there exist ${\varphi _0^{\rm P}}$ and {\color{black} ${\alpha _0^{\rm P}}$} satisfying  $\theta_2 + \theta_3 - \theta_1 + {\varphi _0^{\rm P}} = 0$ and  $\left| {{h_1}} \right| = {\alpha _0^{\rm P }}\left| {{h_2}} \right|\left| {{h_3}} \right| $, and prove that the rate gain of the PT under the above assumption is lower than zero.
Based on this assumption, according to  \eqref{10}, $R^{\rm{P}}_{\rm s}$ can be rewritten as
\begin{align}
{R_{\rm s}^{\rm{P}}} =\sum\limits_{m = 1}^M {\frac{1}{M}{{\log }_2}\left( {1 + \frac{{2P{{\left| {{h_1}} \right|}^2}\left( {1 + \cos \left( {\frac{{2\pi }}{M}\left( {m - 1} \right)}  \right)} \right)}}{{{\sigma ^2}}}} \right)}. \tag{C.1}
\end{align}

Then, we can write  $\Delta {R_s}^{\rm{P}}$ as
\begin{align} \label{c2}\notag
\Delta {R_{\rm s}^{\rm{P}}}&=\sum\limits_{m = 1}^M {\frac{1}{M}{{\log }_2}\left( {1 + 2b\left( {1 +\cos \left( {\frac{{2\pi }}{M}\left( {m - 1} \right)}  \right)} \right)} \!\right)}\\
&- {R_p}, \tag{C.2}
\end{align}
where $b = \frac{{P{{| {{h_1}} |}^2}}}{{{\sigma ^2}}}$.

We rewrite \eqref{c2}  as
\begin{align}
\Delta {R_{\rm s}^{\rm{P}}}= {\log _2}\left( {\frac{{\prod\limits_{m = 1}^M {{{\left( {1 + 2b\left( {1 + \cos \left( {\frac{{2\pi }}{M}\left( {m - 1} \right)} \right)} \right)} \right)}^{\frac{1}{M}}}} }}{{1 + b}}} \right). \tag{C.3}
\end{align}

Due to the uniform distribution and symmetry of the MPSK phase over the interval $\left [ 0,2\pi \right ] $, there exists a $m_0$ such that ${\cos \left( {\frac{{2\pi }}{M}\left( {m_0 - 1} \right)} \right)}=-1$. In this case, we have
\begin{align}
\Delta {R_{\rm s}^{\rm{P}}}={\log _2}\left( {\frac{{\prod\limits_{m = 1,m \ne {m_0}}^M {{{\left( {1 + 2b\left( {1 + \cos \left( {\frac{{2\pi }}{M}\left( {m - 1} \right)} \right)} \right)} \right)}^{\frac{1}{M}}}} }}{{1 + b}}} \right) \tag{C.4}.
\end{align}

Next, to prove $\Delta {R_{\rm s}^{\rm{P}}}<0$,  it is sufficient to show that there exists a possibility for the following inequality to hold, i.e.,  ${\frac{{\prod\limits_{m = 1,m \ne {m_0}}^M {{{\left( {1 + 2b\left( {1 + \cos \left( {\frac{{2\pi }}{M}\left( {m - 1} \right)} \right)} \right)} \right)}^{\frac{1}{M}}}} }}{{1 + b}}}<1$. To this end, we define a function $f\left( b \right)$, given by
\begin{align}
f\left( b \right) = \frac{{\prod\limits_{m = 1,m \ne {m_0}}^M {\left( {1 + 2b\left( {1 + \cos \left( {\frac{{2\pi }}{M}\left( {m - 1} \right)} \right)} \right)} \right)} }}{{{{\left( {1 + b} \right)}^M}}}. \tag{C.5}
\end{align}

It is evident that the sign of  $\Delta {R_{\rm s}^{\rm{P}}}$ is the same as that of $f\left( b \right)$.  As $b$ is sufficiently large, the leading order of $b$ in the numerator of $f\left( b \right)$ is $M-1$, while  in the denominator, it is $M$.
Through asymptotic analysis, it follows that $0<f\left( b \right)<1$, which consequently implies  $\Delta {R_s^{\rm{P}}}<0$.

\subsubsection{Example 2} In this example, we assume that there exist ${\varphi _0^{\rm P}}$ and $\alpha _0^{\rm P}$ meeting $\theta_2 + \theta_3 - \theta_1 + {\varphi _0^{\rm P}} = \frac{\pi}{2}$ and $\left| {{h_1}} \right| = {\alpha _0^{\rm{P}}}\left| {{h_2}} \right|\left| {{h_3}} \right| $, and prove that the rate gain of the PT under the above assumption is larger than zero. $\Delta {R_{\rm s}^{\rm{P}}}$ can be rewritten as
\begin{align}\label{c5}
\Delta {R_{\rm s}^{\rm{P}}}=&\sum\limits_{m = 1}^M {\frac{1}{M}{{\log }_2}\left( {1 + \frac{{2P{{\left| {{h_1}} \right|}^2}}}{{{\sigma ^2}}}} \right)}\nonumber\\
&- {\log _2}\left( {1 + \frac{{P|{h_1}{|^2}}}{{{\sigma ^2}}}} \right) > 0 .\tag{C.6}
\end{align}

\section*{Appendix D}
Based on \eqref{13}, the rate gain of PT can be expressed as
\begin{align}\label{d1}
\Delta {R_{\rm s}^{\rm{P}}}={\log _2}\left( {\frac{{{d_1} + \sqrt {{d_1}^2 - {d_2}^2} }}{2}} \right) - R_p. \tag{D.1}
\end{align}

Through mathematical operations, we have
\begin{align}\label{d2}
&\left(\sqrt {{d_1}^2 - {d_2}^2}\right)^2-\left( 2\left( {1 + \frac{{P\left|{h_1}\right|^2}}{{{\sigma ^2}}}} \right) - {d_1} \right)^2\nonumber\\
&=4\frac{P}{{{\sigma ^2}}}|{h_2}{h_3}{\alpha _0^{\rm{P}}}{|^2}>0,\nonumber\\
 &\Rightarrow \left( {\frac{{{d_1} + \sqrt {{d_1}^2 - {d_2}^2} }}{2}} \right)> {1 + \frac{{P\left|{h_1}\right|^2}}{\sigma ^2}}. \tag{D.2}
\end{align}
Therefore, $\Delta {R_{\rm s}^{\rm{P}}}>0$, indicating that when $M$ is sufficiently large, BD's access leads to an increase in the PT's rate.

\section*{Appendix E}
%

By substituting $y_{\rm{MRC}}=r{e^{j\phi }}$, and defining ${\phi ^\prime } = \phi  - {\varphi _i}$, \eqref{14} can be rewritten as \eqref{f1}, as shown at the bottom of the next page.

%

\begin{figure*}[ht] 
 	\centering
    \begin{align}\label{f1}
    R_{\rm {BD}}^{\rm {P}}=\frac{1}{{M\pi \sigma _s^2}}\sum\limits_{m = 1}^M \int\limits_0^{2\pi } \int\limits_0^{ + \infty } {{\exp\left(^{ - \frac{{{r^2} + {{g }^2}{\alpha _0^{\rm{P}}}^2 - 2r g {\alpha _0^{\rm{P}}}\cos {\phi ^\prime }}}{{\sigma _s^2}}}\right)}}{{\log }_2}\left( {\frac{{\exp \left( {\frac{{2r g {\alpha _0^{\rm{P}}}\cos {\phi ^\prime }}}{{\sigma _s^2}}} \right)}}{{\sum\limits_{i = 1}^M {\exp \left( {\frac{{2rg {\alpha _0^{\rm{P}}}\cos \left( {{\phi ^\prime } - \left( {{\varphi _i} - {\varphi _m}} \right)}\right)}}{{\sigma _s^2}}} \right)} }}} \right) rdrd{\phi ^\prime }+\log_2M. \tag{E.1}
    \end{align}
    \hrulefill
\end{figure*}

For any $m$,  it holds that ${\varphi _i} - {\varphi _m} = \frac{{2\pi }}{M}\left( {i - m} \right)$.
Based on this, by analyzing \eqref{f1}, it can be observed that $R_{\rm {BD}}^{\rm {P}}$ is independent of the phase $\varphi _0^{\rm P}$, but is related to the phase difference between the symbols.


\section*{Appendix F}
Let $A = 1 + \frac{{P\left( {{{\left| {{h_1}} \right|}^2} + {{\left| {{h_2}{h_3}} \right|}^2}{\alpha _0^{\rm{P}}}^2} \right)}}{{{\sigma ^2}}}$, $B = \frac{{2P\left| {{h_1}} \right|\left| {{h_2}} \right|\left| {{h_3}} \right|{\alpha _0^{\rm{P}}}}}{{{\sigma ^2}}}$.
When $M=2$, $R_{\rm s}^{\rm{P}}$ can be written as
\begin{align}\label{g1}
R_{\rm s}^{\rm{P}} &= \sum\limits_{m = 1}^2 {\frac{1}{2}{{\log }_2}\left( {A + B\cos \left( {{\theta _0} + {\varphi _0^{\rm P}} + \pi\left( {m - 1} \right)} \right)} \right)} \nonumber \\
&= \frac{1}{2}{\log _2}\left( {{A^2} - {B^2}{{\cos }^2}\left( {{\theta _0} + {\varphi _0^{\rm P}}} \right)} \right).\tag{F.1}
\end{align}

For $R_{\rm s}^{\rm{P}}$ to be maximized, the condition ${\cos \left({\theta _0}+{\varphi _0^{\rm P}}\right)}=0$ must hold, implying ${\theta _0}+{\varphi _0^{\rm P}} = \frac{\pi }{2} + \eta \pi $, where $\eta \in \mathbb{Z} $.
Specifically, when $M=2$, the phase $\varphi _0^{{\rm P}}$ must satisfy the condition ${\varphi _0^{\rm P}} = \frac{\pi }{2} + \eta \pi  - {\theta _0}$ for maximization of $R_{\rm s}^{\rm{P}}$ under $M=2$.

When $M=4$, $R_{\rm s}^{\rm{P}}$ can be written as
\begin{align}\label{g2}
&{R_{\rm s}^{\rm{P}}} = \sum\limits_{m = 1}^4 {\frac{1}{4}{{\log }_2}\left( {A + B\cos \left( {{\theta _0} + {\varphi _0^{\rm P}} + \frac{\pi }{2}\left( {m - 1} \right)} \right)} \right)} \nonumber \\
&= \frac{1}{4}{\log _2}\left( {{A^4} - {A^2}{B^2} + \frac{{{B^4}}}{4}{{\sin }^2}\left( {2\left( {{\theta _0} + {\varphi _0^{\rm P}}} \right)} \right)} \right).\tag{F.2}
\end{align}

To maximize $R_{\rm s}^{\rm{P}}$, the condition ${{{\sin }^2}\left( {2\left( {{\theta _0} + {\varphi _0^{\rm P}}} \right)} \right)}=1$ must be satisfied,  which implies that $2\left( {{\theta _0} + {\varphi _0^{\rm P}}} \right) = \frac{\pi }{2} + \eta \pi$.
Therefore, for $M=4$, the  phase $\varphi _0^{\rm P}$ should satisfy ${\varphi _0^{\rm P}} = \frac{\pi }{4} + \frac{{\eta \pi }}{2} - {\theta _0}$ to maximize $R_{\rm s}^{\rm{P}}$ under $M=4$.

When $M=2^k$, $R_{\rm s}^{\rm{P}}$ can be written as
\begin{align}\label{g3}
&{R_{\rm s}^{\rm{P}}}=\sum\limits_{m = 1}^{{2^k}} {\frac{1}{{{2^k}}}{{\log }_2}\!\left(\! {A \!+ \!B\cos \left( {{\theta _0} \!+ \!{\varphi _0^{\rm P}} \!+ \!\frac{{2\pi }}{{{2^k}}}\left( {m - 1} \right)} \right)} \right)}\nonumber \\
&= \frac{1}{{{2^k}}}{\log _2}\!\left( \! {\coprod\limits_{m = 1}^{{2^k}} {A \!+\! B\cos \!\left(\! {{\theta _0} \!+\! {\varphi _0^{\rm P}}\! + \! \frac{{2\pi }}{{{2^k}}}\!\left( \!{m - 1} \!\right)}\! \right)} } \right)
.\tag{F.3}
\end{align}

Based on the conditions for maximizing $R_{\rm s}^{\rm{P}}$ for $M=2$ and $M=4$, it is reasonable to assume that for $M=2^k$, the phase $\varphi _0^{\rm P}$ must satisfy ${\varphi _0^{\rm P}} = \frac{\pi }{2^k} + \frac{{2\eta \pi }}{2^k} - {\theta _0}$.
Next, we need to prove whether the phase satisfies ${\varphi _0^{\rm P}} = \frac{\pi }{{{2^{k + 1}}}} + \frac{{2\eta\pi }}{{{2^{k + 1}}}} - {\theta _0}$ holds for $M=2^{k+1}$ when $R_{\rm s}^{\rm{P}}$ is maximized.
If true, the hypothesis is validated.
The detailed process is as follows.

When $M=2^{k+1}$, $R_{\rm s}^{\rm{P}}$ can be written as
\begin{align}\label{g4}
{R_{\rm s}^{\rm{P}}} &\!= \!\sum\limits_{m = 1}^{{2^{k + 1}}} {\frac{1}{{{2^{k + 1}}}}{{\log }_2}\!\left( \!{A \!+ \!B\cos \!\left(\! {{\theta _0} \!+ \!{\varphi _0^{\rm P}} \!+ \!\frac{{2\pi }}{{{2^{k \!+ \! 1}}}}\!\left(\! {m \!- \!1} \!\right)} \!\right)} \! \right)\!} \nonumber \\
&= \frac{1}{{{2^{k + 1}}}}{\log _2}D, \tag{F.4}
\end{align}
where
\begin{align}\label{g5}
D = \prod\limits_{m = 1}^{{2^{k + 1}}} {\left( {A + B\cos \left( {{\theta _0} + {\varphi _0^{\rm P}} + \frac{{2\pi }}{{{2^{k + 1}}}}\left( {m - 1} \right)} \right)} \right)}. \tag{F.5}
\end{align}

Then, \eqref{g5} can be further expanded as
\begin{align} \notag \label{g6}
&D= \underbrace {\prod\limits_{m = 1}^{{2^k}} {\left( {A + B\cos \left( {{\theta _0} + {\varphi _0^{\rm P}} + \frac{{2\pi }}{{{2^{k + 1}}}}\left( {m - 1} \right)} \right)} \right)} }_{{D_1}} \\
&\times \!\underbrace {\prod\limits_{m^{\prime} = {2^k} + 1}^{{2^{k + 1}}} {\!\left(\! {A\! +\! B\cos \!\left(\! {{\theta _0} \!+\! {\varphi _0^{\rm P}}\! +\! \frac{{2\pi }}{{{2^{k + 1}}}}\!\left(\! {m^{\prime} \!- \! 1} \!\right)\!} \right)\!} \right)\!} }_{{D_2}}. \tag{F.6}
\end{align}

Let $m^{\prime \prime}=m^{\prime}-2^k$, $D_2$ can be calculated as
\begin{align}\label{g7}
D_2^{\prime} \!= \!\prod\limits_{{m^{\prime \prime }} = 1}^{{2^k}} {\!\left( \!{A \!-\! B\cos \!\left(\! {{\theta _0}\! + \!{\varphi _0^{\rm P}}\! +\! \frac{{2\pi }}{{{2^{k + 1}}}} \times \!\left(\! {{m^{\prime \prime }} \!-\! 1}\! \right)}\! \right)} \!\right)}. \tag{F.7}
\end{align}

Combining \eqref{g4}, \eqref{g6}, and \eqref{g7}, and let $A^{\prime}={{A^2} - \frac{{{B^2}}}{2}}$, $B^{\prime}={ - \frac{{{B^2}}}{2}}$, $R_{\rm s}^{\rm{P}}$ can be written as \eqref{g8}, as shown at the bottom of the next page.
\begin{figure*}[ht] 
 	\centering
    \begin{align}\label{g8}
    {R_{\rm s}^{\rm{P}}} \!= \!\frac{1}{{{2^{k \!+\! 1}}}}{\log _2}\!\left( \!{\prod\limits_{m \!= \! 1}^{{2^k}} {\!\left( \!{{A^2} \!\!- \!\!{B^2}{{\cos }^2}\!\left( \!{{\theta _0} \!+ \!{\varphi _0^{\rm P}}\! +\! \frac{{2\pi }}{{{2^{k \!+ \!1}}}}\!\left( \!{m \!-\! 1} \!\right)\!} \!\right)\!} \right)\!} } \right)\!= \!\frac{1}{{{2^{k + 1}}}}{\log _2}\!\left( \! {\prod\limits_{m = 1}^{{2^k}} {\!\left( \!{{A^\prime } \!+ \!{B^\prime }\cos \!\left( \!{2\!\left(\! {{\theta _0} \!+\! {\varphi _0^{\rm P}}} \!\right)\! +\! \frac{{2\pi }}{{{2^k}}}  \!\left(\! {m - 1} \!\right) \!} \right)\!} \right)\!} } \right). \tag{F.8}
    \end{align}
    \hrulefill
\end{figure*}
 	
According to \eqref{g3}, we assume that for $M=2^k$, the  phase satisfies ${\varphi _0^{\rm P}} = \frac{\pi }{2^k} + \frac{{2\eta \pi }}{2^k} - {\theta _0}$ when $R_s^{\rm P}$ is maximized, which implies that ${\varphi _0^{\rm P}}+ {\theta _0}= \frac{\pi }{2^k} + \frac{{2\eta \pi }}{2^k}$.
Therefore, from \eqref{g8}, the phase condition becomes $2\left( {{\theta _0} + {\varphi _0^{\rm P}}} \right) = \frac{\pi }{{{2^k}}} + \frac{{2\eta \pi }}{{{2^k}}}$, which simplifies to ${\varphi _0^{\rm P}} = \frac{\pi }{{{2^{k + 1}}}} + \frac{{2\eta\pi }}{{{2^{k + 1}}}} - {\theta _0}$ for $M=2^{k+1}$.

Since this condition holds, the hypothesis is proven, i.e., for $M=2^k$, when $R_{\rm s}^{\rm{P}}$ is maximized, the phase must satisfy ${\varphi _0^{\rm P}} = \frac{\pi }{2^k} + \frac{{2\eta \pi }}{2^k} - {\theta _0}$.

\ifCLASSOPTIONcaptionsoff
  \newpage
\fi
\bibliographystyle{IEEEtran}
\bibliography{refa}

\begin{thebibliography}{10}
\providecommand{\url}[1]{#1}
\csname url@samestyle\endcsname
\providecommand{\newblock}{\relax}
\providecommand{\bibinfo}[2]{#2}
\providecommand{\BIBentrySTDinterwordspacing}{\spaceskip=0pt\relax}
\providecommand{\BIBentryALTinterwordstretchfactor}{4}
\providecommand{\BIBentryALTinterwordspacing}{\spaceskip=\fontdimen2\font plus
\BIBentryALTinterwordstretchfactor\fontdimen3\font minus
  \fontdimen4\font\relax}
\providecommand{\BIBforeignlanguage}[2]{{%
\expandafter\ifx\csname l@#1\endcsname\relax
\typeout{** WARNING: IEEEtran.bst: No hyphenation pattern has been}%
\typeout{** loaded for the language `#1'. Using the pattern for}%
\typeout{** the default language instead.}%
\else
\language=\csname l@#1\endcsname
\fi
#2}}
\providecommand{\BIBdecl}{\relax}
\BIBdecl

\bibitem{8879484}
L.~Chettri and R.~Bera, ``A comprehensive survey on internet of things ({IoT})
  toward {5G} wireless systems,'' \emph{IEEE Internet Things J.}, vol.~7,
  no.~1, pp. 16--32, 2020.

\bibitem{10463656}
M.~M. Butt, N.~R. Mangalvedhe, N.~K. Pratas, J.~Harrebek, J.~Kimionis,
  M.~Tayyab, O.-E. Barbu, R.~Ratasuk, and B.~Vejlgaard, ``Ambient iot: A
  missing link in {3GPP IoT} devices landscape,'' \emph{IEEE Internet Things
  Mag.}, vol.~7, no.~2, pp. 85--92, 2024.

\bibitem{11015785}
S.~Moloudi, T.~A. Khan, G.~Moschetti, R.~Narayanan, H.~Khan, A.~Haskou,
  J.~Bergman, B.~A. Mouris, A.~Höglund, C.~Zhang, D.~Hui, and M.~Afshang,
  ``Ambient power-enabled internet of things: {3GPP} physical layer
  standardization overview,'' \emph{IEEE Commun. Standards Mag.}, vol.~9,
  no.~4, pp. 183--191, 2025.

\bibitem{9749195}
Y.-C. Liang, R.~Long, Q.~Zhang, and D.~Niyato, ``Symbiotic communications:
  Where marconi meets darwin,'' \emph{IEEE Wireless Commun.}, vol.~29, no.~1,
  pp. 144--150, 2022.

\bibitem{xu2025revolutionizing}
R.~Xu, Y.~Ye, H.~Sun, L.~Shi, and G.~Lu, ``Revolutionizing symbiotic radio:
  Exploiting trade-offs in hybrid active-passive communications,'' \emph{IEEE
  Commun. Mag.}, pp. 1--8, 2025.

\bibitem{9193946}
Y.-C. Liang, Q.~Zhang, E.~G. Larsson, and G.~Y. Li, ``Symbiotic radio:
  Cognitive backscattering communications for future wireless networks,''
  \emph{IEEE Trans. Cognit. Commun. Netw.}, vol.~6, no.~4, pp. 1242--1255, Dec.
  2020.

\bibitem{9051982}
Y.~Ye, L.~Shi, X.~Chu, and G.~Lu, ``On the outage performance of ambient
  backscatter communications,'' \emph{IEEE Internet Things J.}, vol.~7, no.~8,
  pp. 7265--7278, 2020.

\bibitem{10980384}
Z.~Cui, G.~Wang, R.~Xu, X.~Wei, F.~Qin, and C.~Tellambura, ``Backscatter
  communications for green internet of things: Practical prototypes, open
  challenges, and standardization,'' \emph{IEEE Internet Things Mag.}, vol.~8,
  no.~3, pp. 32--39, 2025.

\bibitem{8907447}
R.~Long, Y.-C. Liang, H.~Guo, G.~Yang, and R.~Zhang, ``Symbiotic radio: A new
  communication paradigm for passive internet of things,'' \emph{IEEE Internet
  Things J.}, vol.~7, no.~2, pp. 1350--1363, Feb. 2020.

\bibitem{10643608}
Y.~Ye, R.~Xu, G.~Chen, D.~Benevides~da Costa, and G.~Lu, ``Qos-guaranteed
  adaptive power reflection coefficient for self-powered cooperative ambient
  backscatter communication,'' \emph{IEEE Wireless Commun. Lett.}, vol.~13,
  no.~10, pp. 2812--2816, 2024.

\bibitem{10964549}
Z.~Wen, H.~Ding, M.~Elkashlan, D.~Li, C.~Yuen, J.~M. Moualeu, M.~Chen, and
  Z.~Liu, ``Cut to the chase: A fast-decoding scheme for symbiotic backscatter
  multi-user {NOMA} systems,'' \emph{IEEE Trans. Wireless Commun.}, pp. 1--1,
  2025.

\bibitem{10502325}
X.~Song, D.~Han, L.~Shi, H.~Sun, and R.~Q. Hu, ``Relay assisted cooperative
  ambient backscatter communication with hybrid long-short packets,''
  \emph{IEEE Trans. Veh. Technol.}, vol.~73, no.~9, pp. 12\,890--12\,903, 2024.

\bibitem{9461158}
H.~Yang, Y.~Ye, K.~Liang, and X.~Chu, ``Energy efficiency maximization for
  symbiotic radio networks with multiple backscatter devices,'' \emph{{IEEE}
  Open J. Commun. Soc.}, vol.~2, pp. 1431--1444, Jun. 2021.

\bibitem{9866050}
Y.~Ye, L.~Shi, X.~Chu, G.~Lu, and S.~Sun, ``Mutualistic cooperative ambient
  backscatter communications under hardware impairments,'' \emph{IEEE Trans.
  Commun.}, vol.~70, no.~11, pp. 7656--7668, Nov 2022.

\bibitem{8665892}
H.~Guo, Y.-C. Liang, R.~Long, S.~Xiao, and Q.~Zhang, ``Resource allocation for
  symbiotic radio system with fading channels,'' \emph{IEEE Access}, vol.~7,
  pp. 34\,333--34\,347, Mar. 2019.

\bibitem{10437703}
H.~Guo, Y.~Ye, H.~Sun, and L.~Shi, ``Resource allocation for mutualistic
  symbiotic radio with hybrid active-passive communications,'' in \emph{Proc.
  {IEEE} GLOBECOM, Kuala Lumpur, Malaysia}, 2023, pp. 4418--4423.

\bibitem{10820118}
S.~Lu, Y.~Ye, H.~Sun, L.~Shi, and R.~Qingyang~Hu, ``Minimizing total
  transmission time in hybrid active-passive mutualistic symbiotic radio,''
  \emph{IEEE Wireless Commun. Lett.}, vol.~14, no.~3, pp. 846--850, 2025.

\bibitem{8807353}
S.~Zhou, W.~Xu, K.~Wang, C.~Pan, M.-S. Alouini, and A.~Nallanathan, ``Ergodic
  rate analysis of cooperative ambient backscatter communication,'' \emph{IEEE
  Wireless Commun. Lett.}, vol.~8, no.~6, pp. 1679--1682, Dec 2019.

\bibitem{8941106}
H.~Ding, D.~B. da~Costa, and J.~Ge, ``Outage analysis for cooperative ambient
  backscatter systems,'' \emph{IEEE Wireless Commun. Lett.}, vol.~9, no.~5, pp.
  601--605, May. 2020.

\bibitem{10702412}
Y.~Liu, Z.~Zhou, Y.~Ye, X.~Li, M.~Geng, and A.~Nallanathan, ``Outage
  performance analysis for mutualistic symbiotic backscatter communication
  systems,'' \emph{IEEE Trans. Veh. Technol.}, vol.~74, no.~2, pp. 3457--3462,
  Feb 2025.

\bibitem{10778600}
Y.~Ye, Y.~Tian, X.~Chu, S.~Sun, and G.~Lu, ``Outage performance of
  relay-assisted mutualistic backscatter communications under energy-causality
  constraint,'' \emph{IEEE Trans. Commun.}, pp. 1--1, 2024.

\bibitem{10896822}
Z.~Liang, S.~Han, and Y.-C. Liang, ``Outage performance analysis for
  {RIS}-enabling full-duplex cellular symbiotic radio network,'' \emph{IEEE
  Trans. Veh. Technol.}, pp. 1--6, 2025.

\bibitem{8006941}
J.~J. Boutros, F.~Jardel, and C.~Méasson, ``Probabilistic shaping and
  non-binary codes,'' in \emph{Proc. IEEE Int. Symp. Inf. Theory (ISIT)}, 2017,
  pp. 2308--2312.

\bibitem{1505023}
G.~De~Vita and G.~Iannaccone, ``Design criteria for the {RF} section of {UHF}
  and microwave passive {RFID} transponders,'' \emph{IEEE Trans. Microwave
  Theory Tech.}, vol.~53, no.~9, pp. 2978--2990, Sep. 2005.

\bibitem{7059230}
N.~Fasarakis-Hilliard, P.~N. Alevizos, and A.~Bletsas, ``Coherent detection and
  channel coding for bistatic scatter radio sensor networking,'' \emph{IEEE
  Trans. Commun.}, vol.~63, no.~5, pp. 1798--1810, May. 2015.

\bibitem{10681516}
A.~C.~Y. Goay, D.~Mishra, and A.~Seneviratne, ``Optimal reflection coefficients
  for {ASK} modulated backscattering from passive tags,'' \emph{IEEE Trans.
  Commun.}, vol.~73, no.~3, pp. 1692--1708, Mar. 2025.

\bibitem{9309091}
J.~Hu, Y.-C. Liang, and Y.~Pei, ``Reconfigurable intelligent surface enhanced
  multi-user {MISO} symbiotic radio system,'' \emph{IEEE Trans. Commun.},
  vol.~69, no.~4, pp. 2359--2371, 2021.

\bibitem{11020593}
H.~Zhou, Y.-C. Liang, C.~Yuen, and G.~Y. Li, ``Cooperative modulation for
  symbiotic radios: Design methodology, challenges, and solutions,'' \emph{IEEE
  Commun. Mag.}, pp. 1--7, 2025.

\bibitem{1233745}
U.~Karthaus and M.~Fischer, ``Fully integrated passive {UHF} {RFID} transponder
  {IC} with 16.7-{$\mu$W} minimum {RF} input power,'' \emph{IEEE J. Solid-State
  Circuits}, vol.~38, no.~10, pp. 1602--1608, 2003.

\bibitem{7820135}
D.~Darsena, G.~Gelli, and F.~Verde, ``Modeling and performance analysis of
  wireless networks with ambient backscatter devices,'' \emph{IEEE Trans.
  Commun.}, vol.~65, no.~4, pp. 1797--1814, 2017.

\bibitem{6153042}
S.~J. Thomas, E.~Wheeler, J.~Teizer, and M.~S. Reynolds, ``Quadrature amplitude
  modulated backscatter in passive and semipassive {UHF} {RFID} systems,''
  \emph{IEEE Trans. Microwave Theory Tech.}, vol.~60, no.~4, pp. 1175--1182,
  2012.

\bibitem{2011Mutual}
L.~Batina, B.~Gierlichs, E.~Prouff, M.~Rivain, F.~X. Standaert, and
  N.~Veyrat-Charvillon, ``Mutual information analysis: a comprehensive study,''
  \emph{J. Cryptol.}, vol.~24, no.~2, pp. 269--291, 2011.

\bibitem{zwillinger2007table}
D.~Zwillinger and A.~Jeffrey, \emph{Table of integrals, series, and
  products}.\hskip 1em plus 0.5em minus 0.4em\relax Elsevier, 2007.

\end{thebibliography}

\end{document}